\documentclass[11pt]{article}
\usepackage{amssymb,amsmath,amsfonts}
\usepackage{graphicx}
\usepackage{graphics}
\usepackage{epsfig}

\makeatletter
\@addtoreset{equation}{section}

\makeatother

\begin{document}

\title{Radiation processes in dielectric cylindrical waveguides}
\author{A. A. Saharian$^{1,2}$\thanks{%
E-mail: saharian@ysu.am},\thinspace\ L. Sh. Grigoryan$^{2}$, H. F.
Khachatryan$^{2}$ \\
%EndAName
\\
\textit{$^1$Institute of Physics, Yerevan State University,}\\
\textit{1 Alex Manoogian Street, 0025 Yerevan, Armenia} \vspace{0.3cm}\\
\textit{$^2$Institute of Applied Problems of Physics NAS RA,}\\
\textit{25 Hrachya Nersissyan Street, 0014 Yerevan, Armenia}}
\maketitle

\begin{abstract}
Dielectric cylindrical waveguides are widely used for confining and guiding
of electromagnetic waves in relatively wide range of frequencies. They have
found numerous technological and scientific applications in
telecommunications, medicine, material science, photonics and quantum
optics. The electromagnetic field Green function is the central object in
investigations of different types of radiation processes in those
structures. In this paper, we review and further develop the recurrence
procedure for evaluating the electromagnetic field Green function in a
medium made of any number of homogeneous cylindrical layers. The general
results are specified for a cylindrical waveguide immersed in a homogeneous
medium. Expressions are provided for all the components of the Green tensor
in both regions inside and outside the cylinder. As an application of the
results for the Green function, we consider the radiation of a charged
particle rotating around a dielectric cylinder. The intensities for all
types of radiation processes are discussed. They include the
synchrotron-Cherenkov radiation at large distances from the cylinder and the
radiation on guided and surface polaritonic modes confined inside or near
the surface of the cylinder. The paper provides explicit formulas for the
electromagnetic fields and the spectral-angular densities of those
radiations. It also includes a numerical and comparative analysis.
\end{abstract}

\bigskip


\bigskip

\section{Introduction}

The interaction of charged particles with a medium can result in various
radiation processes. One important type of this phenomenon is polarization
radiation, which occurs when the medium is polarized by the particle's
electromagnetic field. In these processes, charged particles act as indirect
sources of radiation; the direct source is the medium's polarization
relaxation. Polarization radiation is exemplified by well-known phenomena
such as Cherenkov radiation \cite{Jell58}-\cite{Afan04}, transition
radiation \cite{Term72}-\cite{Rull98}, and diffraction radiation \cite%
{Poty11,Douc25}. Interest in these radiation processes stems from their
ability to act as controllable sources of radiation across a broad spectrum
of frequencies. These processes are also widely used to investigate the
electromagnetic characteristics of media, for particle beam diagnostics, and
in particle detectors.

Tuning the parameters of radiation sources, the electrodynamic
characteristics of the media, and the geometry of their interfaces can
effectively control the characteristics of generated radiation and its
subsequent propagation. Recent technological advances in designing
metamaterials, photonic crystals, and nanoscale structures offer new
opportunities for manipulating the electric and magnetic characteristics of
materials. In particular, the possibility for the control of dispersion
relations for dielectric permittivity and permeability is of great
importance \cite{Marq08,Tong18}. Significant progress in
metamaterial-related research has led to the stimulation of active
theoretical and experimental investigations of radiation processes in those
media. Of particular interest is the interaction of charged particles with
media whose permittivity and permeability become negative in certain
spectral ranges.

Cylindrical waveguides are essential components used to guide
electromagnetic waves at microwave, millimeter wave, and optical frequencies
in modern communication systems and integrated photonic devices, including
optical amplifiers, modulators, and lasers \cite{Yeh08,Atak13}. They also
play an important role in various fields of fundamental physics, including
light-matter interactions, radiation mechanisms, dispersion relations of
materials, boundary effects, particle beam manipulation and diagnostics, and
dielectric wakefield acceleration. The wide range of applications of
cylindrical waveguides motivates the investigation of different types of
radiation processes in these structures. A great deal of research has been
dedicated to the topic of energy losses of charged particles when they
interact with cylindrical guiding structures (see, e.g., \cite{Bolo62}-\cite%
{Saha25} and references therein). In these papers, different types of
cylindrically symmetric dielectric structures inside a waveguide are
discussed as well. Depending on the geometry of these structures, the
generated radiation is a superposition of Cherenkov, transition, and
diffraction radiations. In the main part of the cited references the charged
particles move along rectilinear trajectories parallel to the axis of the
waveguide. More complicated problems with charges rotating inside or outside
a cylindrical waveguide were considered in references \cite%
{Grig95,Kota00,Kota01,Kota02,Kota02b,Saha05,Saha12,Kota18,Kota19}. The
features of the radiation in the case of helical motion are discussed in 
\cite{Kota07,Saha07,Saha09}. In these problems the radiation field is a
superposition of Cherenkov and synchrotron radiations. Already for a charge
rotating in a homogeneous medium, this superposition gives interesting
interference features \cite{Zrel70},\cite{Kita60}-\cite{Boni86}. The
analysis of radiation processes in more complicated cylindrically symmetric
geometries of waveguides, including those with conical elements, and the
corresponding approximate analytic and numerical methods for study of
radiation features can be found in \cite{Abaj10},\cite{Garc02}-\cite{Grig26}%
.\ 

The present article aims to accomplish several objectives. In the first
part, we review and further develop the recurrence procedure described in 
\cite{Grig95} for evaluating the electromagnetic field Green function in a
medium consisting of general number of homogeneous cylindrical layers. Here,
we provide explicit expressions for the coefficients of the radial parts of
the Green function components in separate layers. The general results are
specified for a cylindrical waveguide immersed in a homogeneous medium. In
this case, the expressions are given for all the components of the Green
tensor in both regions inside and outside the cylinder. This procedure has
been used in a number of papers for the investigation of radiation from
charged particles interacting with a dielectric cylinder. In the previously
discussed problems, depending on the law of particle motion, the separate
components required in those problems were considered. Knowledge of all the
components makes it possible to study radiation processes in the case of a
general law of particle motion. In the second part of the paper, as an
application of the results for the Green function, we consider the radiation
of a charged particle rotating around a dielectric cylinder. The intensities
for all types of radiation processes are discussed. They include the
synchrotron-Cherenkov radiation at large distances from the cylinder, the
radiations on guided and surface polaritonic modes of the cylinder. Explicit
formulas for the electromagnetic fields and the spectral-angular densities
of those radiations are provided. This part of the paper combines some
results from our previous studies.

The organization of the paper is as follows. In the next section, we
describe the electromagnetic field Green function and its relations to the
field potentials in an inhomogeneous medium with cylindrically symmetric
dielectric function. In Section \ref{sec:RecRel}, a recurrence scheme is
developed for the evaluation of the Green function in the special case of
medium consisting general number of homogeneous cylindrical layers. The
construction of the so-called "potential-free" Green function, required at
the initial stage of the recurrent procedure, is presented in Section \ref%
{sec:G0}. In Section \ref{sec:GFwaveguide}, the general scheme is specified
for a homogeneous dielectric cylindrical waveguide immersed in a homogeneous
medium. All the components of the Green function are presented inside and
outside the waveguide. The general features of the radiation processes in
that setup are discussed in Section \ref{sec:Features}. Different types of
the radiation modes and their dispersion relations are considered and the
possibility for the appearance of strong narrow peaks in the angular
distribution of the radiation propagating in the exterior medium is
emphasized. As an radiation source, Section \ref{sec:Circ} considers a point
charge coaxially circulating around a dielectric cylinder. The expressions
for the Fourier components of the scalar and vector potentials, and for the
electric and magnetic fields are presented in both interior and exterior
regions. The radiation propagating in the exterior medium, at large
distances from the cylinder, is discussed in Section \ref{sec: RadInf}. It
presents the interference of the synchrotron and Cherenkov radiations
influenced by the dielectric cylinder. A formula is derived for the
spectral-angular density of the radiation intensity and the presence of
strong narrow peaks in the angular distribution of the radiation on a given
harmonic is demonstrated. In Section \ref{sec:RadGuid}, we consider the
radiation on normal modes of a cylindrical waveguide. The corresponding
contributions to the electric and magnetic fields originate from the poles
of the Green function. Explicit expressions for those parts in the fields
and for energy fluxes through a plane, perpendicular to the cylinder axis,
are provided. The energy fluxes and the radiated power for surface
polaritons are studied in Section \ref{sec:SP}. The features of the
distribution of the radiated normal modes are discussed and numerical
examples are presented. Section \ref{sec:Conc} summarizes the main results
of the paper.

\section{Electromagnetic Green function in cylindrically symmetric media}

\label{sec:GenGD}

The Green function (GF) plays a central role in the study of classical and
quantum electrodynamic effects in media. We consider the electromagnetic
field generated by the current density $\mathbf{j}(x)$ and charge density $%
\rho (x)$ in a cylindrically symmetric nonmagnetic medium. An isotropic and
linear medium will be supposed with the standard relation $\mathbf{D}(x)=%
\hat{\varepsilon}\,\mathbf{E}(x)$ between the electric displacement $\mathbf{%
D}(x)$ and electric field $\mathbf{E}(x)$. Here and below, $x=(t,\mathbf{r})$
stands for spacetime coordinates and the hat on the letter will be used for
operators and matrices. For a medium without spatial dispersion, the action
of the dielectric permittivity operator $\hat{\varepsilon}$ is defined by
the relation $\mathbf{D}(\omega ,\mathbf{r})=\varepsilon (\omega ,\mathbf{r}%
)\,\mathbf{E}(\omega ,\mathbf{r})$ in terms of the spectral components of
the fields, where $\varepsilon (\omega ,\mathbf{r})$ is the
frequency-dependent dielectric permittivity. The spectral component of a
function $f(x)$ is given by $f(\omega ,\mathbf{r})=\int_{-\infty }^{+\infty
}dt\,f(x)e^{i\omega t}/(2\pi )$. In the Lorentz gauge 
\begin{equation}
\mathbf{\nabla }\cdot \mathbf{A}+\frac{\hat{\varepsilon}}{c}\partial
_{t}\varphi =0,  \label{Lor}
\end{equation}%
for the scalar and vector potentials $\varphi (x)$ and $\mathbf{A}(x)$, the
Maxwell's equations read%
\begin{align}
\left( \Delta -\frac{\hat{\varepsilon}}{c^{2}}\partial _{t}^{2}\right) 
\mathbf{A}-\frac{\mathbf{\nabla }\hat{\varepsilon}}{\hat{\varepsilon}}%
\mathbf{\nabla }\cdot \mathbf{A}& =-\frac{4\pi }{c}\mathbf{j},  \label{EqA}
\\
\left( \Delta -\frac{\hat{\varepsilon}}{c^{2}}\partial _{t}^{2}\right)
\varphi +\frac{\mathbf{\nabla }\hat{\varepsilon}}{\hat{\varepsilon}}\cdot
\left( \mathbf{\nabla }\varphi +\frac{1}{c}\partial _{t}\mathbf{A}\right) &
=-4\pi \frac{\rho }{\hat{\varepsilon}}.  \label{Eqxi}
\end{align}%
These equations are obtained from the Maxwell equations in terms of the
vectors $\mathbf{D}(x)$, $\mathbf{E}(x)$, and $\mathbf{B}(x)$, with $\mathbf{%
B}(x)$ being the magnetic field, using the gauge condition (\ref{Lor}) (see
Appendix \ref{sec:App}). The charge and current densities are related by the
continuity equation $\partial _{t}\rho +\mathbf{\nabla }\cdot \mathbf{j}=0$.

Introducing the cylindrical coordinate system $(r,\phi ,z)$, the equation (%
\ref{EqA}) for the vector potential in a cylindrically symmetric medium with 
$\hat{\varepsilon}=\hat{\varepsilon}\left( r\right) $ is presented in the
form%
\begin{equation}
\left( \mathcal{F}_{il}-\frac{\partial _{r}\hat{\varepsilon}}{\hat{%
\varepsilon}}\mathcal{D}_{il}\right) A_{l}=-\frac{4\pi }{c}j_{i},
\label{EqA2}
\end{equation}%
where the indices $i,l=1,2,3$ correspond to the components along the
coordinates $r,\phi ,z$ and a summation is understood over repeated indices.
In (\ref{EqA2}), the matrix operators $\mathcal{\hat{F}}$ and $\mathcal{\hat{%
D}}$ with the matrix elements $\mathcal{F}_{il}$ and $\mathcal{D}_{il}$ are
given by the expressions%
\begin{equation}
\mathcal{\hat{F}}=\left( 
\begin{array}{ccc}
\Delta -r^{-2}-\frac{\hat{\varepsilon}}{c^{2}}\partial _{t}^{2} & 
-2r^{-2}\partial _{\phi } & 0 \\ 
2r^{-2}\partial _{\phi } & \Delta -r^{-2}-\frac{\hat{\varepsilon}}{c^{2}}%
\partial _{t}^{2} & 0 \\ 
0 & 0 & \Delta -\frac{\hat{\varepsilon}}{c^{2}}\partial _{t}^{2}%
\end{array}%
\right) ,  \label{Fcal}
\end{equation}%
and 
\begin{equation}
\mathcal{\hat{D}}=\left( 
\begin{array}{ccc}
1/r+\partial _{r} & r^{-1}\partial _{\phi } & \partial _{z} \\ 
0 & 0 & 0 \\ 
0 & 0 & 0%
\end{array}%
\right) ,  \label{Dcal}
\end{equation}%
where $\Delta $ is the Laplace operator in cylindrical coordinates.

We introduce the electromagnetic field retarded GF $G_{il}(x,x^{\prime })$
related to the operator in the right-hand side of (\ref{EqA2}). Introducing
the $3\times 3$ matrix $\hat{G}(x,x^{\prime })$, with the elements $%
G_{il}(x,x^{\prime })$, the equation for the GF reads 
\begin{equation}
\left( \mathcal{\hat{F}}-\frac{\partial _{r}\hat{\varepsilon}}{\hat{%
\varepsilon}}\mathcal{\hat{D}}\right) G(x,x^{\prime })=\left( 2\pi \right)
^{3}\hat{I}\,\delta \left( x-x^{\prime }\right) ,  \label{GDeq}
\end{equation}%
where $\hat{I}$ is the $3\times 3$ unit matrix. Given the GF, the solution
of the equation (\ref{EqA2}) is presented in the form%
\begin{equation}
A_{i}(x)=-\frac{1}{2\pi ^{2}c}\int d^{4}x^{\prime }\,G_{il}(x,x^{\prime
})j_{l}\left( x^{\prime }\right) .  \label{AiGD}
\end{equation}%
For static and cylindrically symmetric medium we use the partial Fourier
transform%
\begin{equation}
\hat{G}(x,x^{\prime })=\sum_{n=-\infty }^{+\infty }\int_{-\infty }^{+\infty
}dk\int_{-\infty }^{+\infty }d\omega \,\hat{G}_{n}\left( k,\omega
,r,r^{\prime }\right) e^{in\left( \phi -\phi ^{\prime }\right)
+ik(z-z^{\prime })-i\omega (t-t^{\prime })},  \label{GDFour}
\end{equation}%
where $x=(t,r,\phi ,z)$ and $x^{\prime }=(t^{\prime },r^{\prime },\phi
^{\prime },z^{\prime })$. The matrix elements of the Fourier image $\hat{G}%
_{n}\left( k,\omega ,r,r^{\prime }\right) $ will be denoted by $%
G_{il,n}\left( k,\omega ,r,r^{\prime }\right) $. From (\ref{GDeq}) we get
the equation for the Fourier image $\hat{G}_{n}\left( k,\omega ,r,r^{\prime
}\right) \equiv \hat{G}_{n}\left( r,r^{\prime }\right) $: 
\begin{equation}
\left[ \hat{F}-\frac{\partial _{r}\varepsilon (r)}{\varepsilon (r)}\hat{D}%
\right] \hat{G}_{n}\left( r,r^{\prime }\right) =\frac{\hat{I}}{r^{\prime }}%
\delta \left( r-r^{\prime }\right) ,  \label{EqGF}
\end{equation}%
with the dielectric permittivity $\varepsilon (r)=\varepsilon \left( \omega
,r\right) $. The matrix operators $\hat{F}$ and $\hat{D}$ are defined by%
\begin{align}
\hat{F}(r)& =\left( 
\begin{array}{ccc}
\hat{f}(r) & -2in/r^{2} & 0 \\ 
2in/r^{2} & \hat{f}(r) & 0 \\ 
0 & 0 & \hat{f}(r)+1/r^{2}%
\end{array}%
\right) ,  \label{Fop} \\
\hat{D}(r)& =\left( 
\begin{array}{ccc}
1/r+\partial _{r} & in/r & ik \\ 
0 & 0 & 0 \\ 
0 & 0 & 0%
\end{array}%
\right) ,  \label{Dop}
\end{align}%
where%
\begin{equation}
\hat{f}(r)=\frac{1}{r}\partial _{r}\left( r\partial _{r}\right) -\frac{%
n^{2}+1}{r^{2}}+\lambda ^{2},  \label{fL}
\end{equation}%
and 
\begin{equation}
\lambda =\frac{\omega }{c}\sqrt{\varepsilon (r)-c^{2}k^{2}/\omega ^{2}}.
\label{lamb}
\end{equation}

Expansions similar to (\ref{GDFour}) can also be written for the
electromagnetic potentials:%
\begin{equation}
\left\{ 
\begin{array}{c}
A_{i}(x) \\ 
\varphi (x)%
\end{array}%
\right\} =\sum_{n=-\infty }^{+\infty }\int_{-\infty }^{+\infty
}dk\int_{-\infty }^{+\infty }d\omega \,\left\{ 
\begin{array}{c}
A_{i,n}(k,\omega ,r) \\ 
\varphi _{n}(k,\omega ,r)%
\end{array}%
\right\} e^{in\phi +ikz-i\omega t}.  \label{AiFour}
\end{equation}%
The relation between the Fourier components of the vector potential and the
current density is obtained from (\ref{GDFour}):%
\begin{equation}
\,A_{i,n}(k,\omega ,r)=-\frac{4\pi }{c}\int_{0}^{\infty }dr^{\prime
}\,r^{\prime }G_{il,n}\left( k,\omega ,r,r^{\prime }\right) j_{l,n}(k,\omega
,r^{\prime }),  \label{AiGDf}
\end{equation}%
where%
\begin{equation}
j_{l,n}(k,\omega ,r)=\frac{1}{\left( 2\pi \right) ^{3}}\int_{0}^{2\pi }d\phi
\int_{-\infty }^{+\infty }dz\int_{-\infty }^{+\infty }dt\,j_{l}(x)e^{-in\phi
-ikz+i\omega t},  \label{jf}
\end{equation}%
is the Fourier image of the current density. Having the vector potential,
the Fourier component of the scalar potential is found from the Lorentz
gauge condition (\ref{Lor}):%
\begin{equation}
\varphi _{n}(k,\omega ,r)=\frac{c}{\omega \varepsilon (r)}\left[ \frac{n}{r}%
A_{2,n}(k,\omega ,r)+kA_{3,n}(k,\omega ,r)-\frac{i}{r}\partial _{r}\left(
rA_{1,n}(k,\omega ,r)\right) \right] .  \label{phif}
\end{equation}%
The strengths of the electric and magnetic fields are determined using the
standard relations with the potentials.

\section{Recurrence scheme for evaluation of the GF in cylindrically
stratified media}

\label{sec:RecRel}

In this section, we will consider a special case of the general setup
presented in the previous section. The medium consists of $N$ homogeneous
cylindrical layers with dielectric permittivities $\varepsilon
_{0},\varepsilon _{1},\ldots ,\varepsilon _{N}$. The radial dependence of
the dielectric function is given by%
\begin{equation}
\varepsilon (r)=\varepsilon \left( \omega ,r\right) =\varepsilon
_{0}+\sum_{s=1}^{N}\left( \varepsilon _{s}-\varepsilon _{s-1}\right) \theta
\left( r-r_{s}\right) ,  \label{epscyl}
\end{equation}%
where $\theta (y)$ is the Heaviside unit step function and $\varepsilon
_{s}=\varepsilon _{s}(\omega )$ for $s=0,1,\ldots ,N$. Substituting this in (%
\ref{EqGF}), we get the equation for the Fourier image of the GF:%
\begin{equation}
\left[ \hat{F}(r)-\sum_{s=1}^{N}\hat{A}^{(s)}(r)\right] \hat{G}_{n}\left(
r,r^{\prime }\right) =\frac{\hat{I}}{r^{\prime }}\delta \left( r-r^{\prime
}\right) ,  \label{EqGF2}
\end{equation}%
where the matrix $\hat{F}$ is given by (\ref{Fop}) with dielectric
permittivity (\ref{epscyl}) in the definition (\ref{lamb}) of $\lambda $,
and 
\begin{equation}
\hat{A}^{(s)}(r)=\frac{\varepsilon _{s}-\varepsilon _{s-1}}{\varepsilon (r)}%
\delta \left( r-r_{s}\right) \hat{D}(r).  \label{Akpot}
\end{equation}%
The terms in (\ref{EqGF2}) with the functions $\hat{A}^{(s)}(r)$ can be
interpreted as delta-type "interaction potentials" localized on the
separating boundaries between homogeneous media. The boundary conditions for
the Green function at the interfaces between the cylindrical layers are
encoded in the equation (\ref{EqGF2}). The Green function is continuous on
the separating boundaries and the conditions for the jump of radial
derivatives are obtained in standard way for equations containing delta-type
potentials, namely, by integrating the equation (\ref{EqGF2}) over a small
interval of radial coordinate $r$ near the boundary $r=r_{s}$. In the
discussion below, for an example of a cylindrical waveguide in a homogeneous
medium, it will be shown that this leads to standard boundary conditions for
the electric and magnetic fields.

We define the $3\times 3$ Green function $\hat{G}_{n}^{(0)}\left(
r,r^{\prime }\right) $, with the elements $G_{il,n}^{(0)}\left( r,r^{\prime
}\right) $, by the equation%
\begin{equation}
\hat{F}(r)G_{n}^{(0)}\left( r,r^{\prime }\right) =\frac{\hat{I}}{r^{\prime }}%
\delta \left( r-r^{\prime }\right) .  \label{GD0eq}
\end{equation}%
For $\varepsilon (r)=\mathrm{const}$ it reduces to the GF for a homogeneous
medium. The equation (\ref{GD0eq}) is obtained from (\ref{EqGF2}) in the
limit of zero delta-type "potentials" (\ref{Akpot}). In this sense, the
function $G_{n}^{(0)}\left( r,r^{\prime }\right) $ can be called a
"potential-free" GF. With this function, the equation (\ref{EqGF2}) can be
rewritten in the form of Lipmann-Schwinger equation%
\begin{equation}
\hat{G}_{n}(r,r^{\prime })=\hat{G}_{n}^{(0)}(r,r^{\prime
})+\sum_{s=1}^{N}\int_{0}^{\infty }dr^{\prime \prime }\,r^{\prime \prime }%
\hat{G}_{n}^{(0)}(r,r^{\prime \prime })\hat{A}^{(s)}(r^{\prime \prime })\hat{%
G}_{n}(r^{\prime \prime },r^{\prime }).  \label{EqGF3}
\end{equation}%
Let us introduce intermediate $3\times 3$ matrices $\hat{G}%
_{n}^{(s)}(r,r^{\prime })$, $s=1,2,\ldots ,N$, defined by the relations%
\begin{equation}
\hat{G}_{n}^{(s)}(r,r^{\prime })=\hat{G}_{n}^{(s-1)}(r,r^{\prime
})+\int_{0}^{\infty }dr^{\prime \prime }\,r^{\prime \prime }\hat{G}%
_{n}^{(s-1)}(r,r^{\prime \prime })\hat{A}^{(s)}(r^{\prime \prime })\hat{G}%
_{n}^{(s)}(r^{\prime \prime },r^{\prime }).  \label{GFs}
\end{equation}%
Writing this equation for $s$ replaced by $s-1$ and putting the
corresponding expression for $\hat{G}_{n}^{(s-1)}(r,r^{\prime })$ in (\ref%
{GFs}), it is seen that%
\begin{equation}
\hat{G}_{n}^{(s)}(r,r^{\prime })=\hat{G}_{n}^{(s-2)}(r,r^{\prime
})+\int_{0}^{\infty }dr^{\prime \prime }\,r^{\prime \prime }\hat{G}%
_{n}^{(s-2)}(r,r^{\prime \prime })\left[ \hat{A}^{(s-1)}(r^{\prime \prime })+%
\hat{A}^{(s)}(r^{\prime \prime })\right] \hat{G}_{n}^{(s)}(r^{\prime \prime
},r^{\prime }).  \label{GFs2}
\end{equation}%
Repeating this procedure, we get%
\begin{equation}
\hat{G}_{n}^{(s)}(r,r^{\prime })=\hat{G}_{n}^{(s-i)}(r,r^{\prime
})+\int_{0}^{\infty }dr^{\prime \prime }\,r^{\prime \prime }\hat{G}%
_{n}^{(s-i)}(r,r^{\prime \prime })\sum_{l=s+1-i}^{s}\hat{A}^{(l)}(r^{\prime
\prime })\hat{G}_{n}^{(s)}(r^{\prime \prime },r^{\prime }).  \label{GFsi}
\end{equation}%
For $s=i=N$, this equation is reduced to (\ref{EqGF3}) and, hence, $\hat{G}%
_{n}^{(N)}(r,r^{\prime })=\hat{G}_{n}(r,r^{\prime })$.

By taking into account the expression (\ref{Akpot}) for $\hat{A}^{(s)}(r)$
in (\ref{GFs}), one finds%
\begin{equation}
G_{n}^{(s)}(r,r^{\prime })=G_{n}^{(s-1)}(r,r^{\prime })+\left( \varepsilon
_{s}-\varepsilon _{s-1}\right) \,r_{s}G_{n}^{(s-1)}(r,r_{s})\frac{\hat{D}%
(r^{\prime \prime })}{\varepsilon (r^{\prime \prime })}G_{n}^{(s)}(r^{\prime
\prime },r^{\prime })|_{r^{\prime \prime }=r_{s}}.  \label{GFs3}
\end{equation}%
Here and below, for a given function $g(r,r^{\prime })$, the substitution $%
r=r_{s}$ is understood in the sense%
\begin{equation}
\frac{\hat{D}(r)}{\varepsilon (r)}g(r,r^{\prime })|_{r=r_{s}}=\frac{\hat{D}%
(r)}{2\varepsilon _{s-1}}g(r,r^{\prime })|_{r=r_{s}-0}+\frac{\hat{D}(r)}{%
2\varepsilon _{s}}g(r,r^{\prime })|_{r=r_{s}+0}.  \label{Dnot}
\end{equation}%
Acting on (\ref{GFs3}) from the left by $(1/\varepsilon (r))\hat{D}(r)$ and
then taking $r=r_{s}$, we obtain 
\begin{equation}
\left[ \hat{I}-\hat{S}_{s}(r_{s},r_{s})\right] \left( \varepsilon
_{s}-\varepsilon _{s-1}\right) \,r_{s}\frac{\hat{D}(r)}{\varepsilon (r)}\hat{%
G}_{n}^{(s)}(r,r^{\prime })|_{r=r_{s}}=\hat{S}_{s}(r_{s},r^{\prime }),
\label{DFs4}
\end{equation}%
where we have introduced the $3\times 3$ matrix 
\begin{equation}
\hat{S}_{s}(r_{s},r^{\prime })=\left( \varepsilon _{s}-\varepsilon
_{s-1}\right) \,r_{s}\frac{\hat{D}(r)}{\varepsilon (r)}\hat{G}%
_{n}^{(s-1)}(r,r^{\prime })|_{r=r_{s}},  \label{Ss}
\end{equation}%
in the sense of (\ref{Dnot}).

The relation (\ref{DFs4}) allows us to find the quantity (\ref{Dnot})
appearing in the right-hand side of (\ref{GFs3}). To do so, note that,
according to the definition (\ref{Dop}) of the matrix operator $\hat{D}(r)$,
for the elements of the matrix $\hat{S}_{s}(r_{s},r^{\prime })$ one has $%
S_{s,il}(r_{s},r^{\prime })=0$ for $i=2,3$. For this type of matrices, the
relation%
\begin{equation}
\left[ \hat{I}-\hat{S}_{s}(r_{s},r^{\prime })\right] ^{-1}=\hat{I}+\frac{%
\hat{S}_{s}(r_{s},r^{\prime })}{1-\mathrm{Sp}\,\hat{S}_{s}(r_{s},r^{\prime })%
},  \label{Srel}
\end{equation}%
holds. By taking into account this relation, from (\ref{DFs4}) we obtain%
\begin{equation}
\left( \varepsilon _{s}-\varepsilon _{s-1}\right) \,r_{s}\frac{\hat{D}(r)}{%
\varepsilon (r)}\hat{G}_{n}^{(s)}(r,r^{\prime })|_{r=r_{s}}=\left[ \hat{I}+%
\frac{\hat{S}_{s}(r_{s},r_{s})}{1-\mathrm{Sp}\,\hat{S}_{s}(r_{s},r_{s})}%
\right] \hat{S}_{s}(r_{s},r^{\prime }).  \label{Rel2}
\end{equation}%
Substituting this in the relation (\ref{GFs3}) one finds%
\begin{equation}
\hat{G}_{n}^{(s)}(r,r^{\prime })=\hat{G}_{n}^{(s-1)}(r,r^{\prime })+\hat{G}%
_{n}^{(s-1)}(r,r_{s})\left[ \hat{I}+\frac{\hat{S}_{s}(r_{s},r_{s})}{1-%
\mathrm{Sp}\,\hat{S}_{s}(r_{s},r_{s})}\right] \hat{S}_{s}(r_{s},r^{\prime }).
\label{Gsrec}
\end{equation}%
This provides the recurrence relation which allows to find the GF $\hat{G}%
_{n}^{(s)}(r,r^{\prime })$ having the function $\hat{G}_{n}^{(s-1)}(r,r^{%
\prime })$. Note that for the elements $S_{s,il}(r_{s},r^{\prime })$ of the
matrix $\hat{S}_{s}(r_{s},r^{\prime })$ we have $S_{s,il}(r_{s},r^{\prime
})=0$ for $i=2,3$ and, hence, $\mathrm{Sp}\,\hat{S}%
_{s}(r_{s},r_{s})=S_{s,11}(r_{s},r^{\prime })$. By taking into account that $%
S_{s,il}(r_{s},r^{\prime })=0$ for $i=2,3$, the relation%
\begin{equation}
\left[ \hat{I}+\frac{\hat{S}_{s}(r_{s},r_{s})}{1-\mathrm{Sp}\,\hat{S}%
_{s}(r_{s},r_{s})}\right] \hat{S}_{s}(r_{s},r^{\prime })=\frac{\hat{S}%
_{s}(r_{s},r^{\prime })}{1-\mathrm{Sp}\,\hat{S}_{s}(r_{s},r_{s})},
\label{RelS}
\end{equation}%
can be proved. With this result, the formula (\ref{Gsrec}) is rewritten in a
simpler form (an alternative representation is given in \cite{Grig95})%
\begin{equation}
\hat{G}_{n}^{(s)}(r,r^{\prime })=\hat{G}_{n}^{(s-1)}(r,r^{\prime })+\frac{%
\hat{G}_{n}^{(s-1)}(r,r_{s})\hat{S}_{s}(r_{s},r^{\prime })}{1-\mathrm{Sp}\,%
\hat{S}_{s}(r_{s},r_{s})},  \label{Gsrec2}
\end{equation}%
for $s=1,2,\ldots ,N$ and $\hat{G}_{n}(r,r^{\prime })=\hat{G}%
_{n}^{(N)}(r,r^{\prime })$. The formula (\ref{Gsrec2}) provides a recurrence
relation for finding the GF $\hat{G}_{n}(r,r^{\prime })$ given the function $%
\hat{G}_{n}^{(0)}(r,r^{\prime })$.

\section{Construction of the function $\hat{G}^{(0)}(r,r^{\prime })$}

\label{sec:G0}

In accordance with the results obtained in the previous section, the
starting point of the determination of the GF is the evaluation of the
function $\hat{G}_{n}^{(0)}(r,r^{\prime })$, which obeys the equation (\ref%
{GD0eq}). To diagonalize this matrix equation, we perform the transformation%
\begin{equation}
\hat{G}_{n}^{(0)\prime }(r,r^{\prime })=\hat{M}^{-1}\hat{G}%
_{n}^{(0)}(r,r^{\prime })\hat{M},\;\hat{F}^{\prime }(r)=\hat{M}^{-1}\hat{F}%
(r)\hat{M},  \label{Trans1}
\end{equation}%
with the matrix%
\begin{equation}
\hat{M}=\left( 
\begin{array}{ccc}
1 & -i\delta _{n} & 0 \\ 
-i\delta _{n} & 1 & 0 \\ 
0 & 0 & 1%
\end{array}%
\right) ,\;\delta _{n}=1-\delta _{n0}.  \label{Mtrans}
\end{equation}%
In the new representation, the equation (\ref{GD0eq}) takes the form%
\begin{equation}
\hat{F}^{\prime }(r)\hat{G}_{n}^{(0)\prime }(r,r^{\prime })=\frac{\hat{I}}{r}%
\delta (r-r^{\prime }),  \label{GD0eq2}
\end{equation}%
with the diagonal matrix operator%
\begin{equation}
\hat{F}^{\prime }(r)=\mathrm{diag}\left( \hat{f}(r)-\frac{2n}{r^{2}},\hat{f}%
(r)+\frac{2n}{r^{2}},\hat{f}(r)+\frac{1}{r^{2}}\right) ,  \label{Fopdiag}
\end{equation}%
and with $\hat{f}(r)$ given by (\ref{fL}).

The solution of the equation (\ref{GD0eq2}) is presented as%
\begin{equation}
\hat{G}_{n}^{(0)\prime }(r,r^{\prime })=\mathrm{diag}\left(
g_{n+1},g_{n-1},g_{n}\right) ,  \label{GD0sol}
\end{equation}%
where the function $g_{n}=g_{n}(r,r^{\prime })$ obeys the equation%
\begin{equation}
\left[ \hat{f}(r)+\frac{1}{r^{2}}\right] g_{n}(r,r^{\prime })=\frac{1}{r}%
\delta (r-r^{\prime }).  \label{Eqgn}
\end{equation}%
The solution for $\hat{G}^{(0)}(r,r^{\prime })$ is obtained by inverting the
transformation (\ref{Trans1}). For $n\neq 0$ we get%
\begin{equation}
\hat{G}_{n}^{(0)}(r,r^{\prime })=\frac{1}{2}\left( 
\begin{array}{ccc}
g_{n+1}+g_{n-1} & i\left( g_{n+1}-g_{n-1}\right) & 0 \\ 
-i\left( g_{n+1}-g_{n-1}\right) & g_{n+1}+g_{n-1} & 0 \\ 
0 & 0 & 2g_{n}%
\end{array}%
\right) ,  \label{GD0sol2}
\end{equation}%
and for $n=0$ one obtains $\hat{G}_{0}^{(0)}(r,r^{\prime })=\mathrm{diag}%
\left( g_{1},g_{1},g_{0}\right) $. Note that the latter is obtained from (%
\ref{GD0sol2}) taking $n=0$ and $g_{-1}=g_{1}$. The functions $g_{\pm
1}(r,r^{\prime })$ are the solutions of the same equation (\ref{Eqgn}).

The result (\ref{GD0sol2}) is valid for general case of cylindrically
symmetric dielectric permittivity $\varepsilon (r)=\varepsilon \left( \omega
,r\right) $, with $\lambda $ in (\ref{fL}) defined by (\ref{lamb}). For a
stratified medium with homogeneous cylindrical layers, the permittivity is
given by (\ref{epscyl}) with constant dielectric functions $\varepsilon
_{i}=\varepsilon _{i}\left( \omega \right) $, $i=0,1,2,\ldots ,N$. In this
case the equation (\ref{Eqgn}) for the function $g_{n}(r,r^{\prime })$ in
separate layers is reduced to the Bessel equation for $r\neq r^{\prime }$.
Let $r_{j}<r^{\prime }<r_{j+1}$. Then, in the layer $r_{i}<r<r_{i+1}$ with $%
i\neq j$, the solution to the equation (\ref{Eqgn}) is a linear combination
of the Bessel function $J_{n}(y)$ and the Hankel function of the first kind $%
H_{n}^{(1)}(y)\equiv H_{n}(y)$:%
\begin{equation}
g_{n}(r,r^{\prime })=c_{i}J_{n}(\lambda _{i}r)+b_{i}H_{n}(\lambda _{i}r),
\label{gni}
\end{equation}%
where 
\begin{equation}
\lambda _{i}=\frac{\omega }{c}\sqrt{\varepsilon _{i}-c^{2}k^{2}/\omega ^{2}}.
\label{lambi}
\end{equation}%
From the equation (\ref{GD0eq}) it follows that the GF $\hat{G}%
_{n}^{(0)}(r,r^{\prime })$ and its first derivative $\partial _{r}\hat{G}%
_{n}^{(0)}(r,r^{\prime })$ are continuous at $r=r_{i}$ and, hence, the same
holds for the functions $g_{n}(r,r^{\prime })$. This gives the conditions%
\begin{equation}
c_{i-1}J_{n}^{(l)}(\lambda _{i-1}r_{i})+b_{i-1}H_{n}^{(l)}(\lambda
_{i-1}r_{i})=c_{i}J_{n}^{(l)}(\lambda _{i}r_{i})+b_{i}H_{n}^{(l)}(\lambda
_{i}r_{i}),  \label{condri}
\end{equation}%
where $l=0,1$ and $h^{(0)}(r)=h(r)$, $h^{(1)}(r)=h^{\prime }(r)$ for a given
function $h(r)$. In the $j$-th layer we have%
\begin{equation}
g_{n}(r,r^{\prime })=\left\{ 
\begin{array}{cc}
\bar{c}_{j}J_{n}(\lambda _{j}r)+\bar{b}_{j}H_{n}(\lambda _{j}r), & 
r_{j}<r<r^{\prime } \\ 
c_{j}J_{n}(\lambda _{j}r)+b_{j}H_{n}(\lambda _{j}r), & r^{\prime }<r<r_{j+1}%
\end{array}%
\right. .  \label{gnj}
\end{equation}%
The matching conditions at $r=r^{\prime }$ read%
\begin{equation}
\bar{c}_{j}J_{n}^{(l)}(\lambda _{j}r^{\prime })+\bar{b}_{j}H_{n}^{(l)}(%
\lambda _{j}r^{\prime })=c_{j}J_{n}^{(l)}(\lambda _{j}r^{\prime
})+b_{j}H_{n}^{(l)}(\lambda _{j}r^{\prime })-\delta _{l1}/r^{\prime },
\label{condrj}
\end{equation}%
again, with $l=0,1$. We have two additional conditions on the coefficients.
The first one is the regularity condition on the axis $z$ and gives $b_{0}=0$
for $r<r_{0},r^{\prime }$. The second condition requires traveling
cylindrical waves propagating to the infinity and requires $c_{N}=0$ for $%
r>r^{\prime },r_{N}$. Hence, the number of the coefficients in the
expressions for the functions $g_{n}(r,r^{\prime })$ is equal to $2N+2$. The
conditions (\ref{condri}) and (\ref{condrj}) provide $2N+2$ equations to
determine those coefficients.

From the conditions (\ref{condri}) and for $i\neq j$ the following
recurrence relations are obtained between the coefficients:%
\begin{align}
c_{i} &=\frac{\pi }{2i}\left(
c_{i-1}V_{(i)n}^{JH}+b_{i-1}V_{(i)n}^{HH}\right) ,  \notag \\
b_{i} &=-\frac{\pi }{2i}\left(
c_{i-1}V_{(i)n}^{JJ}+b_{i-1}V_{(i)n}^{HJ}\right) ,  \label{cbrec}
\end{align}%
where we have defined 
\begin{equation}
V_{(i)n}^{FG}=F_{n}(\lambda _{i-1}r_{i})r_{i}\partial _{r_{i}}G_{n}(\lambda
_{i}r_{i})-G_{n}(\lambda _{i}r_{i})r_{i}\partial _{r_{i}}F_{n}(\lambda
_{i-1}r_{i}),  \label{VFG}
\end{equation}%
for given functions $F_{n}(y)$ and $G_{n}(y)$. By using the recurrence
relations for the Bessel functions, we also have%
\begin{equation}
V_{(i)n}^{FG}=pr_{i}\left[ \lambda _{i-1}F_{n+p}(\lambda
_{i-1}r_{i})G_{n}(\lambda _{i}r_{i})-\lambda _{i}F_{n}(\lambda
_{i-1}r_{i})G_{n+p}(\lambda _{i}r_{i})\right] ,  \label{VFG2}
\end{equation}%
for $p=\pm 1$. From the boundary condition at $r=r_{j}$ one finds%
\begin{align}
\bar{c}_{j} &=\frac{\pi }{2i}\left(
c_{j-1}V_{(j)n}^{JH}+b_{j-1}V_{(j)n}^{HH}\right) ,  \notag \\
\bar{b}_{j} &=\frac{\pi i}{2}\left(
c_{j-1}V_{(j)n}^{JJ}+b_{j-1}V_{(j)n}^{HJ}\right) .  \label{cbbar}
\end{align}%
From (\ref{condrj}), the relations 
\begin{equation}
c_{j}=\bar{c}_{j}+\frac{i\pi }{2}H_{n}(\lambda _{j}r^{\prime }),\;b_{j}=\bar{%
b}_{j}-\frac{i\pi }{2}J_{n}(\lambda _{j}r^{\prime }),  \label{relcbj}
\end{equation}%
are obtained for the coefficients in the region $r_{j}<r<r_{j+1}$. Note that
for $i\neq j$ we can also obtain the recurrence relations of the form%
\begin{align}
c_{i-1} &=\frac{\pi i}{2}\left[ c_{i}V_{(i)n}^{HJ}+b_{i}V_{(i)n}^{HH}\right]
,  \notag \\
b_{i-1} &=\frac{\pi }{2i}\left[ c_{i}V_{(i)n}^{JJ}+b_{i}V_{(i)n}^{JH}\right]
.  \label{cbrec2}
\end{align}%
Combining the formulas (\ref{cbrec}) and (\ref{cbrec2}), the relation 
\begin{equation}
V_{(i)n}^{HJ}V_{(i)n}^{JH}-V_{(i)n}^{JJ}V_{(i)n}^{HH}=\frac{4}{\pi ^{2}}
\label{relVHG}
\end{equation}%
is obtained for the functions (\ref{VFG}).

For $r_{j}<r^{\prime }<r_{j+1}$ and $j\geq 1$, from the regularity condition
in the region $r<r_{1}$ we have $b_{0}=0$. From the relations (\ref{cbrec})
and (\ref{cbbar}) it follows that all the coefficients $c_{i}$ and $b_{i}$
with $i<j$ and $\bar{c}_{j}$ and $\bar{b}_{j}$ are determined by $c_{0}$: $%
c_{i}=C_{i}^{\mathrm{(in)}}c_{0}$, $b_{i}=B_{i}^{\mathrm{(in)}}c_{0}$, $\bar{%
c}_{i}=\bar{C}_{i}^{\mathrm{(in)}}c_{0}$, and $\bar{b}_{i}=\bar{B}_{i}^{%
\mathrm{(in)}}c_{0}$. Here, the coefficients $C_{i}^{\mathrm{(in)}}$ and $%
B_{i}^{\mathrm{(in)}}$ are uniquely determined in terms of the functions (%
\ref{VFG}). In a similar manner, from the condition of traveling wave in the
region $r>r_{N}$ we have $c_{N}=0$ and, from the relations (\ref{cbrec2}),
all the coefficients $c_{i}$ and $b_{i}$ with $i\geq j$ are expressed in
terms of $b_{N}$ as $c_{i}=C_{i}^{\mathrm{(out)}}b_{N}$ and $b_{i}=B_{i}^{%
\mathrm{(out)}}b_{N}$. The coefficients $c_{0}$ and $b_{N}$ are determined
from the relations (\ref{relcbj}).

\section{GF for a cylindrical waveguide immersed in a homogeneous medium}

\label{sec:GFwaveguide}

Now, let us specify the general procedure described above for the simple
case of an inhomogeneous medium with a single boundary ($N=1$) at $r=r_{1}$.
This corresponds to a waveguide with a dielectric permittivity $\varepsilon
_{0}(\omega )$ that is immersed in a homogeneous medium with a permittivity $%
\varepsilon _{1}(\omega )$. Two cases should be distinguished. For $%
0<r^{\prime }<r_{1}$, in the region $r<r_{1}$ we have $\bar{b}_{0}=0$, and $%
b_{0}$ is determined from (\ref{relcbj}) with $j=0$. In the region $r>r_{1}$
one has the function (\ref{gni}) with $i=1$ and $c_{1}=0$. $c_{0}$ is
determined from the first relation (\ref{cbrec}) with $i=1$, and $b_{1}$ is
determined from the second relation in (\ref{cbrec}). The expression for $%
b_{1}$ is further simplified by using the relation (\ref{relVHG}). The final
expression for the function $g_{n}(r,r^{\prime })$ in the case $r^{\prime
}<r_{1}$ reads%
\begin{equation}
g_{n}(r,r^{\prime })=\left\{ 
\begin{array}{cc}
\frac{i\pi }{2}\left[ \frac{V_{n}^{HH}}{V_{n}^{JH}}J_{n}(\lambda
_{0}r_{>})-H_{n}(\lambda _{0}r_{>})\right] J_{n}(\lambda _{0}r_{<}), & 
r<r_{1} \\ 
\frac{J_{n}(\lambda _{0}r^{\prime })}{V_{n}^{JH}}H_{n}(\lambda _{1}r), & 
r>r_{1}%
\end{array}%
\right. ,  \label{gnCyl1}
\end{equation}%
where $r_{>}=\mathrm{max}(r,r^{\prime })$ and $r_{<}=\mathrm{min}%
(r,r^{\prime })$, and we used the simplified notation $V_{n}^{FG}\equiv
V_{(1)n}^{FG}$ with $V_{(1)n}^{FG}$ defined by (\ref{VFG}). For the region $%
r^{\prime }<r_{1}$ one has $b_{0}=c_{1}=0$ and from the relations given
above we get 
\begin{equation}
g_{n}(r,r^{\prime })=\left\{ 
\begin{array}{cc}
\frac{H_{n}(\lambda _{1}r^{\prime })}{V_{n}^{JH}}J_{n}(\lambda _{0}r), & 
r<r_{1} \\ 
\frac{i\pi }{2}\left[ \frac{V_{n}^{JJ}}{V_{n}^{JH}}H_{n}(\lambda
_{1}r_{<})-J_{n}(\lambda _{1}r_{<})\right] H_{n}(\lambda _{1}r_{>}), & 
r>r_{1}%
\end{array}%
\right. .\;  \label{gncyl2}
\end{equation}%
The expressions for the functions $g_{n\pm 1}(r,r^{\prime })$ are obtained
from (\ref{gnCyl1}) and (\ref{gncyl2}) by the replacements $n\rightarrow
n\pm 1$.

With the functions (\ref{gnCyl1}) and (\ref{gncyl2}), the GF $\hat{G}%
_{n}^{(0)}(r,r^{\prime })$ is given by (\ref{GD0sol2}). The GF for a
dielectric cylinder is obtained from the general formula (\ref{Gsrec2}) with 
$s=1$ and $\hat{G}_{n}^{(1)}(r,r^{\prime })=\hat{G}_{n}(r,r^{\prime })$:%
\begin{equation}
\hat{G}_{n}(r,r^{\prime })=\hat{G}_{n}^{(0)}(r,r^{\prime })+\frac{\hat{G}%
_{n}^{(0)}(r,r_{1})\hat{S}(r_{1},r^{\prime })}{1-\mathrm{Sp}\,\hat{S}%
(r_{1},r_{1})},  \label{Gncyl}
\end{equation}%
where $\hat{S}(r_{1},r^{\prime })=\hat{S}_{1}(r_{1},r^{\prime })$ and%
\begin{equation}
\hat{S}(r_{1},r^{\prime })=r_{1}\frac{\varepsilon _{1}^{2}-\varepsilon
_{0}^{2}}{2\varepsilon _{0}\varepsilon _{1}}\,\hat{D}(r)\hat{G}%
_{n}^{(0)}(r,r^{\prime })|_{r=r_{1}}.  \label{Sr1}
\end{equation}%
From the structure of the operator $\hat{D}(r)$, for the matrix elements of
the product $\hat{D}(r)\hat{G}_{n}^{(0)}(r,r^{\prime })$ we have $(\hat{D}(r)%
\hat{G}_{n}^{(0)}(r,r^{\prime }))_{il}=0$ for $i=2,3$. The nonzero elements
of the first row in the region $r^{\prime }>r_{1}$ are expressed as%
\begin{equation}
\left( \hat{D}(r)\hat{G}_{n}^{(0)}(r,r^{\prime })\right) _{1l}|_{r=r_{1}}=%
\frac{1}{2}J_{n}(\lambda _{0}r_{1})\left( \lambda _{0}\sum_{p}\frac{%
pH_{n+p}(\lambda _{1}r^{\prime })}{V_{n+p}^{JH}},i\lambda _{0}\sum_{p}\frac{%
H_{n+p}(\lambda _{1}r^{\prime })}{V_{n+p}^{JH}},2ik\frac{H_{n}(\lambda
_{1}r^{\prime })}{V_{n}^{JH}}\right) ,  \label{DrG0}
\end{equation}%
and for $r^{\prime }<r_{1}$ we have%
\begin{equation}
\left( \hat{D}(r)\hat{G}_{n}^{(0)}(r,r^{\prime })\right) _{1l}|_{r=r_{1}}=%
\frac{1}{2}H_{n}(\lambda _{1}r_{1})\left( \lambda _{1}\sum_{p}\frac{%
pJ_{n+p}(\lambda _{0}r^{\prime })}{V_{n+p}^{JH}},i\lambda _{1}\sum_{p}\frac{%
J_{n+p}(\lambda _{0}r^{\prime })}{V_{n+p}^{JH}},2ik\frac{J_{n}(\lambda
_{0}r^{\prime })}{V_{n}^{JH}}\right) ,  \label{Drg01}
\end{equation}%
where $l=1,2,3$ and $\sum_{p}=\sum_{p=\pm 1}$. By using (\ref{VFG2}), the
representation 
\begin{equation}
V_{n+p}^{JH}=pr_{1}\left[ \lambda _{1}J_{n+p}(\lambda
_{0}r_{1})H_{n}(\lambda _{1}r_{1})-\lambda _{0}J_{n}(\lambda
_{0}r_{1})H_{n+p}(\lambda _{1}r_{1})\right] ,  \label{VJHp}
\end{equation}%
is obtained for the combination of the Bessel and Hankel functions in (\ref%
{DrG0}) and (\ref{Drg01}).

By taking into account that $\mathrm{Sp}\,\hat{S}%
(r_{1},r_{1})=S_{11}(r_{1},r_{1})$, we get%
\begin{equation}
1-\mathrm{Sp}\,\hat{S}(r_{1},r_{1})=\frac{\varepsilon _{1}^{2}-\varepsilon
_{0}^{2}}{2\varepsilon _{0}\varepsilon _{1}}\alpha _{n}\left( \lambda
_{01},\lambda _{11}\right) ,  \label{SpS}
\end{equation}%
where the function%
\begin{equation}
\alpha _{n}\left( k\right) =\frac{\varepsilon _{0}}{\varepsilon
_{1}-\varepsilon _{0}}-\frac{\lambda _{01}}{2}J_{n}(\lambda
_{01})\sum_{l=\pm 1}l\frac{H_{n+l}(\lambda _{11})}{V_{n+l}^{JH}}
\label{alfn}
\end{equation}%
is introduced with the notations%
\begin{equation}
\lambda _{i1}=\lambda _{i}r_{1},\;i=0,1.  \label{lami1}
\end{equation}%
With the expressions given above, the GF is obtained from (\ref{Gncyl}).

Let us introduce the function $G^{\mathrm{(c)}}(r,r^{\prime })$ in
accordance with%
\begin{equation}
\hat{G}^{\mathrm{(c)}}(r,r^{\prime })=\left\{ 
\begin{array}{ll}
\hat{G}(r,r^{\prime }), & \left( r-r_{1}\right) \left( r^{\prime
}-r_{1}\right) <0 \\ 
\hat{G}(r,r^{\prime })-\hat{G}_{\varepsilon _{i}}(r,r^{\prime }), & \left(
r-r_{1}\right) \left( r^{\prime }-r_{1}\right) >0%
\end{array}%
\right. ,  \label{Gc}
\end{equation}%
where $G_{\varepsilon _{i}}(r,r^{\prime })$ is the GF for a homogeneous
medium with permittivity $\varepsilon _{i}$. The latter is given by 
\begin{equation}
\hat{G}_{\varepsilon _{i}}(r,r^{\prime })=\frac{1}{2}\left( 
\begin{array}{ccc}
\sum_{p}g_{n+p}^{(0)} & i\sum_{p}pg_{n+p}^{(0)} & 0 \\ 
-i\sum_{p}pg_{n+p}^{(0)} & \sum_{p}g_{n+p}^{(0)} & 0 \\ 
0 & 0 & 2g_{n}^{(0)}%
\end{array}%
\right) ,  \label{Ghom}
\end{equation}%
with the function%
\begin{equation}
g_{n}^{(0)}=g_{n}^{(0)}(\lambda _{i}r,\lambda _{i}r^{\prime })=\frac{\pi }{2i%
}J_{n}(\lambda _{i}r_{<})H_{n}(\lambda _{i}r_{>}).  \label{gn0}
\end{equation}%
where, as before, $r_{>}=\mathrm{max}(r,r^{\prime })$ and $r_{<}=\mathrm{min}%
(r,r^{\prime })$. In (\ref{Gc}), $\varepsilon _{i}=\varepsilon _{0}$ for $%
r<r_{1}$ and $\varepsilon _{i}=\varepsilon _{1}$ for $r>r_{1}$. For $%
r,r^{\prime }>r_{1}$ we have $\hat{G}^{\mathrm{(c)}}(r,r^{\prime })=\hat{G}%
(r,r^{\prime })-\hat{G}_{\varepsilon _{1}}(r,r^{\prime })$ and $\hat{G}^{%
\mathrm{(c)}}(r,r^{\prime })$ presents the contribution to the GF in the
exterior region due to the presence of a dielectric cylinder with
permittivity $\varepsilon _{0}$. Similarly, for $r,r^{\prime }<r_{1}$, the
function $\hat{G}^{\mathrm{(c)}}(r,r^{\prime })$ in the interior regions is
interpreted as the part of the GF which comes from the replacement $%
\varepsilon _{0}\rightarrow \varepsilon _{1}$ in the region $r>r_{1}$. It
can be seen that, similar to the function (\ref{Ghom}), one has%
\begin{equation}
G_{3l}(r,r^{\prime })=0,\;l=1,2,  \label{G3l0}
\end{equation}%
in both exterior and interior regions. We give the expressions for the
nonzero components of the GF $\hat{G}^{\mathrm{(c)}}(r,r^{\prime })$ inside
and outside the waveguide separately.

\subsection{GF inside the cylinder}

In the region inside the cylinder, $r<r_{1}$, we obtain 
\begin{equation}
G_{lm}^{\mathrm{(c)}}(r,r^{\prime })=\frac{\left( -i\right) ^{\delta _{m3}}}{%
2}\sum_{p}\left( ip\right) ^{m-l}C_{n}^{(mp)}\frac{J_{n+p}(\lambda _{0}r)}{%
V_{n+p}^{JH}},  \label{Glmi}
\end{equation}%
for $l=1,2$ and $m=1,2,3$. Here, the coefficients are defined by%
\begin{align}
C_{n}^{(mp)} &=H_{n+p}(\lambda _{1}r^{\prime })+\frac{\lambda
_{01}J_{n}(\lambda _{01})}{2p^{m-1}\alpha _{n}\left( k\right) }%
H_{n+p}(\lambda _{11})\sum_{l=\pm 1}\frac{H_{n+l}(\lambda _{1}r^{\prime })}{%
l^{m}V_{n+l}^{JH}},\;r^{\prime }>r_{1},  \notag \\
C_{n}^{(mp)} &=\frac{i\pi }{2}V_{n+p}^{HH}J_{n+p}(\lambda _{0}r^{\prime })+%
\frac{\lambda _{11}H_{n}(\lambda _{11})}{2p^{m-1}\alpha _{n}\left( k\right) }%
H_{n+p}(\lambda _{11})\sum_{l=\pm 1}\frac{J_{n+l}(\lambda _{0}r^{\prime })}{%
l^{m}V_{n+l}^{JH}},\;r^{\prime }<r_{1},  \label{Cni}
\end{align}%
for $m=1,2$, and%
\begin{equation}
C_{n}^{(3p)}=\frac{kr_{1}H_{n+p}(\lambda _{11})}{\alpha _{n}\left( k\right)
V_{n}^{JH}}J_{n}(\lambda _{0}r_{1<}^{\prime })H_{n}(\lambda
_{1}r_{1>}^{\prime }),  \label{Cn3}
\end{equation}%
where $r_{1>}^{\prime }=\mathrm{max}(r_{1},r^{\prime })$ and $r_{1<}^{\prime
}=\mathrm{min}(r_{1},r^{\prime })$. For the remaining nonzero components one
gets%
\begin{align}
G_{33}^{\mathrm{(c)}}(r,r^{\prime }) &=H_{n}(\lambda _{1}r^{\prime })\frac{%
J_{n}(\lambda _{0}r)}{V_{n}^{JH}},\;r^{\prime }>r_{1},  \notag \\
G_{33}^{\mathrm{(c)}}(r,r^{\prime }) &=\frac{i\pi }{2}\frac{V_{n}^{HH}}{%
V_{n}^{JH}}J_{n}(\lambda _{0}r^{\prime })J_{n}(\lambda _{0}r),\;r^{\prime
}<r_{1}.  \label{G33i}
\end{align}%
In these expressions,%
\begin{equation}
V_{n}^{FG}=\lambda _{11}F_{n}(\lambda _{01})G_{n}^{\prime }(\lambda
_{11})-\lambda _{01}G_{n}(\lambda _{11})F_{n}^{\prime }(\lambda _{01}),
\label{VnFG}
\end{equation}%
with $F,G=J,H$ and $V_{n+l}^{JH}$ is given by (\ref{VJHp}).

\subsection{GF outside the cylinder}

Outside the cylinder, $r>r_{1}$, the components with $l=1,2$ and $m=1,2,3$
are expressed as 
\begin{equation}
G_{lm}^{\mathrm{(c)}}(r,r^{\prime })=\frac{\left( -i\right) ^{\delta _{m3}}}{%
2}\sum_{p}\left( ip\right) ^{m-l}C_{n}^{(mp)}\frac{H_{n+p}(\lambda _{1}r)}{%
V_{n+p}^{JH}},  \label{Glme}
\end{equation}%
with the coefficients%
\begin{align}
C_{n}^{(mp)} &=\frac{i\pi }{2}V_{n+p}^{JJ}H_{n+p}(\lambda _{1}r^{\prime })+%
\frac{\lambda _{01}J_{n}(\lambda _{01})}{2p^{m-1}\alpha _{n}\left( k\right) }%
J_{n+p}(\lambda _{01})\sum_{l=\pm 1}\frac{H_{n+l}(\lambda _{1}r^{\prime })}{%
l^{m}V_{n+l}^{JH}},\;r^{\prime }>r_{1},  \notag \\
C_{n}^{(mp)} &=J_{n+p}(\lambda _{0}r^{\prime })+\frac{\lambda
_{11}H_{n}(\lambda _{11})}{2p^{m-1}\alpha _{n}\left( k\right) }%
J_{n+p}(\lambda _{01})\sum_{l=\pm 1}\frac{J_{n+l}(\lambda _{0}r^{\prime })}{%
l^{m}V_{n+l}^{JH}},\;r^{\prime }<r_{1},  \label{Cnme}
\end{align}%
for $m=1,2$, and%
\begin{equation}
C_{n}^{(3p)}=\frac{kr_{1}J_{n+p}(\lambda _{01})}{\alpha _{n}\left( k\right)
V_{n}^{JH}}J_{n}(\lambda _{0}r_{1<}^{\prime })H_{n}(\lambda
_{1}r_{1>}^{\prime }).  \label{Cn3e}
\end{equation}%
The 33-component is given by the expressions%
\begin{align}
G_{33}^{\mathrm{(c)}}(r,r^{\prime }) &=\frac{i\pi }{2}\frac{V_{n}^{JJ}}{%
V_{n}^{JH}}H_{n}(\lambda _{1}r^{\prime })H_{n}(\lambda _{1}r),\;r^{\prime
}>r_{1},  \notag \\
G_{33}^{\mathrm{(c)}}(r,r^{\prime }) &=\frac{J_{n}(\lambda _{0}r^{\prime })}{%
V_{n}^{JH}}H_{n}(\lambda _{1}r),\;r^{\prime }<r_{1},  \label{G33e}
\end{align}%
with the definition (\ref{VnFG}).

The formulas given above for the components of the GF apply to the general
case of dielectric permittivities in separate media. In particular, the
given expressions can be used to obtain the corresponding results for
waveguides with conducting walls. For example, taking the formal limit $%
\varepsilon _{N}\rightarrow \infty $, we get the expressions for the
electromagnetic GF inside a conducting cylindrical waveguide with radius $%
r_{N}$ filled with cylindrically layered dielectric medium with
permittivities $\varepsilon _{0},\varepsilon _{1},\ldots ,\varepsilon _{N-1}$%
.

\section{General features of radiation processes}

\label{sec:Features}

Having the GF, the Fourier components of the vector and scalar potentials
are found by using the relations (\ref{AiGDf}) and (\ref{phif}). Then,
expanding the strengths $\mathbf{E}(x)$ and $\mathbf{B}(x)$ of the electric
and magnetic fields%
\begin{equation}
\left\{ 
\begin{array}{c}
\mathbf{E}(x) \\ 
\mathbf{B}(x)%
\end{array}%
\right\} =\sum_{n=-\infty }^{+\infty }\int_{-\infty }^{+\infty
}dk\int_{-\infty }^{+\infty }d\omega \,\left\{ 
\begin{array}{c}
\mathbf{E}_{n}(k,\omega ,r) \\ 
\mathbf{B}_{n}(k,\omega ,r)%
\end{array}%
\right\} e^{in\phi +ikz-i\omega t},  \label{EHexp}
\end{equation}%
the Fourier components are obtained from the standard formulas $\mathbf{H}%
(x)=\mathbf{\nabla }\times \,\mathbf{A}(x)$ and $\mathbf{E}(x)=-(1/c)\mathbf{%
\partial }_{t}\,\mathbf{A}(x)-\mathbf{\nabla }\varphi $. The obtained
expressions for the electromagnetic fields are valid for general case of
complex dielectric permittivities $\varepsilon _{i}(\omega )$.

In the idealized case of real functions $\varepsilon _{i}(\omega )$, the
Fourier components of the GF may have poles for real values of $k$ (on the
integration axis in the Fourier expansion). These poles correspond to the
normal modes (eigenmodes) of the dielectric cylinder and they are roots of
the equation%
\begin{equation}
\alpha _{n}\left( k\right) =0.  \label{alfn0}
\end{equation}%
By taking into account that 
\begin{equation}
V_{n}^{JH}=\lambda _{11}J_{n}(\lambda _{01})H_{n}^{\prime }(\lambda
_{11})-\lambda _{01}H_{n}(\lambda _{11})J_{n}^{\prime }(\lambda _{01}),
\label{VJH}
\end{equation}%
the equation (\ref{alfn0}) is written as%
\begin{equation}
\left( \lambda _{0}\frac{H_{n}^{\prime }}{H_{n}}-\lambda _{1}\frac{%
J_{n}^{\prime }}{J_{n}}\right) \left( \varepsilon _{1}\lambda _{0}\frac{%
H_{n}^{\prime }}{H_{n}}-\varepsilon _{0}\lambda _{1}\frac{J_{n}^{\prime }}{%
J_{n}}\right) =\frac{n^{2}k^{2}\omega _{n}^{2}}{c^{2}r_{1}^{2}\lambda
_{1}^{2}\lambda _{0}^{2}}\left( \varepsilon _{1}-\varepsilon _{0}\right)
^{2},  \label{NormMode}
\end{equation}%
where $H_{n}=H_{n}(\lambda _{11})$, $J_{n}=J_{n}(\lambda _{01})$, and the
prime stands for the derivative of the function with respect to the
argument. For real $\varepsilon _{i}$, the quantities $\lambda _{i1}^{2}$
are real and the right-hand side of (\ref{NormMode}) is real. The equation (%
\ref{NormMode}) determines the dispersion relation $\omega =\omega (k)$ for
waves with a given $n$. It can be shown that for $\lambda _{1}^{2}>0$ the
equation (\ref{NormMode}) has no solutions. The radial dependence of the
corresponding Fourier modes is given in terms of the Hankel functions $%
H_{n}(\lambda _{1}r)$ and they describe the radiation propagating in the
exterior medium $r>r_{1}$ at large distances from the waveguide.

For modes with $\lambda _{1}^{2}<0$, the radial dependence is given in terms
of the modified Bessel function $K_{n}(\gamma _{1}r)$ and the corresponding
fields exponentially decay at large distances from the cylinder. Here and
below, we use the notations%
\begin{equation}
\gamma _{i}^{2}=k^{2}-\omega ^{2}\varepsilon _{i}/c^{2},\;\gamma
_{i1}=\gamma _{i}r_{1},\;i=0,1.  \label{gami}
\end{equation}%
In this case, the equation for the eigenmodes (\ref{NormMode}) is written as
(see, for example, \cite{Jack99}),%
\begin{equation}
U_{n}(k)\equiv V_{n}\left( \varepsilon _{1}\lambda _{01}\frac{K_{n}^{\prime }%
}{K_{n}}+\varepsilon _{0}\gamma _{11}\frac{J_{n}^{\prime }}{J_{n}}\right) -%
\frac{n^{2}k^{2}\omega _{n}^{2}}{c^{2}\gamma _{1}^{2}\lambda _{0}^{2}}\left(
\varepsilon _{1}-\varepsilon _{0}\right) ^{2}=0,  \label{NormMode2}
\end{equation}%
where $K_{n}=K_{n}\left( \gamma _{11}\right) $ and the notation%
\begin{equation}
V_{n}=\lambda _{01}\frac{K_{n}^{\prime }}{K_{n}}+\gamma _{11}\frac{%
J_{n}^{\prime }}{J_{n}}  \label{Vn}
\end{equation}%
is introduced. With these notations, the function (\ref{alfn}) is presented
in the form 
\begin{equation}
\alpha _{n}(k)=\frac{U_{n}(k)}{(\varepsilon _{1}-\varepsilon _{0})\left(
V_{n}^{2}-n^{2}u^{2}\right) },\;u=\frac{\lambda _{0}}{\gamma _{1}}+\frac{%
\gamma _{1}}{\lambda _{0}}.  \label{alfn1}
\end{equation}%
The equation (\ref{NormMode2}) describes two classes of modes. The first one
corresponds to $\lambda _{0}^{2}>0$ and the radial dependence inside the
cylinder is expressed in terms of the Bessel function $J_{n}(\lambda _{0}r)$%
. These modes present the guided modes of a dielectric cylinder.

For the second class of modes one has $\lambda _{0}^{2}<0$ and the radial
dependence inside the cylinder is described by the modified Bessel function $%
I_{n}(\gamma _{0}r)$. These modes are localized near the surface of the
waveguide and correspond to the surface polaritonic degrees of freedom. An
important example of surface polaritons are surface plasmon polaritons \cite%
{Maie07,Enoc12,Stoc18}. They present collective oscillations of electron
subsystem coupled to electromagnetic field and propagating along an
interface between two media. Due to important properties, such as high
density of electromagnetic energy and the possibility of concentrating
electromagnetic fields beyond the diffraction limit of light waves, surface
polaritons have found applications in a wide range of fundamental and
applied fields. The latter include plasmonic waveguides, light-emitting
devices, surface imaging, plasmonic solar cells, etc.

The equation for surface polaritonic eigenmodes, localized near a
cylindrical surface, is obtained from (\ref{NormMode2}) by passing to the
modified Bessel function (see also \cite{Ashl74,Khos91}):%
\begin{equation}
\left( \frac{I_{n}^{\prime }}{\gamma _{0}I_{n}}-\frac{K_{n}^{\prime }}{%
\gamma _{1}K_{n}}\right) \left( \frac{\varepsilon _{0}I_{n}^{\prime }}{%
\gamma _{0}I_{n}}-\frac{\varepsilon _{1}K_{n}^{\prime }}{\gamma _{1}K_{n}}%
\right) =\frac{n^{2}k^{2}\omega _{n}^{2}}{r_{1}^{2}c^{2}\gamma
_{1}^{4}\gamma _{0}^{4}}\left( \varepsilon _{1}-\varepsilon _{0}\right) ^{2},
\label{NormMode3}
\end{equation}%
with $I_{n}=I_{n}(\gamma _{01})$. By taking into account that $I_{n}^{\prime
}/I_{n}>n/\gamma _{01}$ and $K_{n}^{\prime }/K_{n}<-n/\gamma _{11}$, it can
be shown that the equation (\ref{NormMode3}) has no solutions if the
dielectric permittivities $\varepsilon _{0}$ and $\varepsilon _{1}$ have the
same sign. This corresponds to the well-known condition for the presence of
a surface polaritonic mode at the interface between two media: the
dielectric permittivities of the neighboring media must have opposite signs
within the considered spectral range.

As mentioned above, for real $\varepsilon _{i}$ and for waves propagating at
large distances from the cylinder (corresponding to $\lambda _{1}^{2}>0$),
the function $\alpha _{n}\left( k\right) $ has no zeros within the
integration range of the Fourier expansions (\ref{EHexp}). Mathematically,
this is related to the fact that for $\lambda _{1}^{2}>0$ this function is a
complex function and it is not possible to have zero real and imaginary
parts simultaneously. However, as discussed in \cite{Kota02b,Saha05} for
special cases of the charge motion, under certain conditions, at the zeros
of the real part of the function $\alpha _{n}\left( k\right) $ the imaginary
part can be exponentially small. This can lead to strong, narrow peaks in
the angular distribution of the radiation intensity at large distances from
the cylinder. We will consider this possibility in detail for the general
case of charge motion.

From Debye's asymptotic expansion for the cylinder functions with large
values of the order, $n\gg 1$, and for $|y|<1$, one has \cite{Abra72}%
\begin{equation}
\frac{J_{n}(ny)}{Y_{n}(ny)}\sim \frac{\mathrm{sgn}\,y}{2}\,e^{-2n\zeta
(y)},\;\zeta (y)=\ln \frac{1+\sqrt{1-y^{2}}}{|y|}-\sqrt{1-y^{2}},  \label{JY}
\end{equation}%
where $Y_{n}(u)$ is the Neumann function. By taking into account that this
ratio is exponentially small, for the function $\alpha _{n}\left( k\right) $
in the range 
\begin{equation}
0<\frac{\omega ^{2}}{c^{2}}\varepsilon _{1}<n^{2}+k^{2}  \label{condp}
\end{equation}%
(this corresponds to the condition $|y|<1$ in (\ref{JY})) we can write%
\begin{equation}
\alpha _{n}\left( k\right) \approx \alpha _{n}^{(0)}\left( \lambda
_{01},\lambda _{11}\right) -\frac{i}{2}\lambda _{01}J_{n}(\lambda
_{01})\sum_{l=\pm 1}l\frac{Y_{n+l}(\lambda _{11})}{V_{n+l}^{JY}}\left[ \frac{%
V_{n+l}^{JJ}}{V_{n+l}^{JY}}-\frac{J_{n+l}(\lambda _{11})}{Y_{n+l}(\lambda
_{11})}\right] ,  \label{alfanExp}
\end{equation}%
where 
\begin{equation}
\alpha _{n}^{(0)}\left( k\right) =\frac{\varepsilon _{0}}{\varepsilon
_{1}-\varepsilon _{0}}-\frac{\lambda _{01}}{2}J_{n}(\lambda
_{01})\sum_{l=\pm 1}l\frac{Y_{n+l}(\lambda _{11})}{V_{n+l}^{JY}}.
\label{alfan0}
\end{equation}%
The function $\alpha _{n}^{(0)}\left( k\right) $ is real and it may have
zeros. At these zeros and for large values of $n$, the function $\alpha
_{n}\left( k\right) $ is approximated by the last term in (\ref{alfanExp})
and it is suppressed by the factor $e^{-2n\zeta (\lambda _{11}/n)}$. This
suggests that the intensity of radiation propagating in the exterior medium
may have strong peaks at the zeros of $\alpha _{n}^{(0)}\left( k\right) $.

To clarify the existence of solutions to the equation $\alpha
_{n}^{(0)}\left( k\right) =0$, it is convenient to present it in the form%
\begin{equation}
\left( \lambda _{0}\frac{Y_{n}^{\prime }}{Y_{n}}-\lambda _{1}\frac{%
J_{n}^{\prime }}{J_{n}}\right) \left( \lambda _{0}\frac{\varepsilon
_{1}Y_{n}^{\prime }}{\varepsilon _{0}Y_{n}}-\lambda _{1}\frac{J_{n}^{\prime }%
}{J_{n}}\right) =\frac{n^{2}}{r_{1}^{2}}\left( 1-\frac{\lambda _{0}^{2}}{%
\lambda _{1}^{2}}\right) \left( \frac{\lambda _{1}^{2}}{\lambda _{0}^{2}}-%
\frac{\varepsilon _{1}}{\varepsilon _{0}}\right) ,  \label{PeakEq}
\end{equation}%
where $Y_{n}=Y_{n}(\lambda _{11})$. This equation is obtained from (\ref%
{NormMode}) by the replacement $H_{n}\rightarrow Y_{n}$. The different form
of the right-hand side is convenient for the analysis. From the uniform
asymptotic expansions of the cylinder functions for large values of the
order $n$ one has%
\begin{equation}
\frac{J_{n}^{\prime }(ny)}{J_{n}(ny)}\sim -\frac{Y_{n}^{\prime }(ny)}{%
Y_{n}(ny)}\sim \frac{\sqrt{1-y^{2}}}{y},  \label{Jas}
\end{equation}%
in the range $0<y<1$, and 
\begin{equation}
\frac{J_{n}^{\prime }(ny)}{J_{n}(ny)}\sim -\tan \left[ n\zeta _{1}(y)-\frac{%
\pi }{4}\right] \frac{\sqrt{y^{2}-1}}{y},  \label{Jas1}
\end{equation}%
for $y>1$, with $\zeta _{1}(y)=\sqrt{y^{2}-1}-\mathrm{arcsec\,}y$. In the
region $\omega ^{2}\varepsilon _{0}<k^{2}c^{2}$, the ratio $J_{n}^{\prime
}/J_{n}$ is replaced by $-iI_{n}^{\prime }/I_{n}$, $I_{n}=I_{n}(\gamma
_{11}) $, and the corresponding leading asymptotic reads%
\begin{equation}
\frac{I_{n}^{\prime }(ny)}{I_{n}(ny)}\sim \frac{\sqrt{1+y^{2}}}{y},\;y>0.
\label{Ias}
\end{equation}%
By using these asymptotics, it can be shown that for large $n$ the equation (%
\ref{PeakEq}) has a solution only under the condition%
\begin{equation}
\frac{\omega ^{2}}{c^{2}}\varepsilon _{0}>\frac{n^{2}}{r_{1}^{2}}+k^{2}.
\label{condP2}
\end{equation}%
Now, combining this with (\ref{condp}), as a necessary condition, we get $%
\varepsilon _{0}>\varepsilon _{1}$. Of course, the existence of solutions to
(\ref{PeakEq}) not necessarily mean that peaks will appear in the radiation
intensity. In addition to the factor $1/\alpha _{n}^{(0)}\left( k\right) $,
other factors in the corresponding formula can offset the contribution of
this part. An example of the appearance of strong peaks will be considered
below.

\section{Electromagnetic fields for a charge circulating around a cylinder}

\label{sec:Circ}

As an application of the general setup described above, we consider the
radiation generated by a charged particle moving with constant velocity $v$
along a circular trajectory of radius $r=r_{0}>r_{1}$ . For the charge and
current densities we have%
\begin{equation}
j_{i}=\rho v_{i}=\frac{\delta _{2i}}{r}qv\delta \left( r-r_{0}\right) \delta
\left( \phi -\omega _{0}t\right) \delta (z),  \label{ji}
\end{equation}%
where $q$ is the charge of the particle and $\omega _{0}=v/r_{0}$ is the
angular velocity of the rotation.

\subsection{Vector and scalar potentials}

Substituting the Fourier component 
\begin{equation}
j_{i,n}(k,\omega ,r)=\frac{\delta _{2i}}{\left( 2\pi \right) ^{2}}\frac{qv}{r%
}\delta (r-r_{0})\delta \left( \omega -n\omega _{0}\right) ,  \label{jiF}
\end{equation}%
in (\ref{AiGDf}), for the vector potential we get%
\begin{equation}
\,A_{i,n}(k,\omega ,r)=-\frac{qv}{\pi c}G_{i2,n}\left( k,\omega
,r,r_{0}\right) \delta \left( \omega -n\omega _{0}\right) .  \label{Aic}
\end{equation}%
In this formula, the delta function is related to the periodic motion of the
radiation source. It is also present in the Fourier components of the scalar
potential and the field strengths. For a field $X(x)$ with the Fourier
component $X_{n}(k,\omega ,r)$, we define $X_{n}(k,r)$ in accordance with $%
X_{n}(k,\omega ,r)=X_{n}(k,r)\delta \left( \omega -n\omega _{0}\right) $.
The presence of the delta function allows to write the Fourier expansions (%
\ref{AiFour}) and (\ref{EHexp}) in the form%
\begin{align}
X(x) &=\sum_{n=-\infty }^{+\infty }\int_{-\infty }^{+\infty
}dk\,X_{n}(k,r)e^{in\left( \phi -\omega _{0}t\right) +ikz}  \notag \\
&={\mathrm{Re}}\left[ \sum_{n=0}^{\infty }\left( 2-\delta _{0n}\right)
\int_{-\infty }^{+\infty }dk\,X_{n}(k,r)e^{in\left( \phi -\omega
_{0}t\right) +ikz}\right] ,  \label{Xexpc}
\end{align}%
for $X=\varphi ,\mathbf{A},\mathbf{E},\mathbf{B}$. For real functions $X(x)$%
, we have the relation $X_{n}^{\ast }(k,r)=X_{-n}(-k,r)$ and we have used it
in the second line. In particular, for the vector potential we have%
\begin{equation}
A_{i,n}(k,r)=-\frac{qv}{\pi c}G_{i2,n}\left( k,n\omega _{0},r,r_{0}\right) .
\label{AinG}
\end{equation}%
By taking into account that $G_{32,n}(k,\omega ,r,r^{\prime })=0$, we
conclude that $A_{3,n}(k,r)=0$. The expressions for nonzero components of
the vector potential are obtained by using the formulas for the components
of the GF given above.

We denote the fields and their Fourier images in the region with dielectric
permittivity $\varepsilon _{j}$, $j=0,1$, by $X(x)=X^{(j)}(x)$ and $%
X_{n}^{(j)}(k,r)$, respectively. With such an arrangement, the functions $%
X_{n}^{(0)}(k,r)$ and $X_{n}^{(1)}(k,r)$ correspond to the fields in the
regions $r<r_{0}$ and $r>r_{0}$. Then, the corresponding vector potentials
are presented in a combined form%
\begin{equation}
A_{m,n}^{(j)}(k,r)=A_{m,n}^{(\varepsilon _{1})}(k,r)\delta _{1j}+\frac{qv}{%
2\pi c}\sum_{p}\frac{C_{(j)n}^{(p)}}{\left( ip\right) ^{m}}%
Z_{n+p}^{(j)}(\lambda _{j}r),  \label{Ajc}
\end{equation}%
where $m=1,2$, 
\begin{equation}
\lambda _{j}^{2}=\frac{\omega _{n}^{2}}{c^{2}}\varepsilon
_{j}-k^{2},\;\omega _{n}=|n|\omega _{0},  \label{lamj2}
\end{equation}%
and we use the notation%
\begin{equation}
Z_{\nu }^{(j)}(y)=\left\{ 
\begin{array}{cc}
J_{\nu }(y), & j=0 \\ 
H_{\nu }(y), & j=1%
\end{array}%
\right. .  \label{Znu}
\end{equation}%
In (\ref{Ajc}), 
\begin{equation}
A_{m,n}^{(\varepsilon _{j})}(k,r)=-\frac{iqv}{4c}\sum_{p}\left( -ip\right)
^{m}J_{n+p}(\lambda _{j}r_{0<})H_{n+p}(\lambda _{j}r_{0>}),  \label{Aihom}
\end{equation}%
with $r_{0>}=\mathrm{max}(r_{0},r)$ and $r_{0<}=\mathrm{min}(r_{0},r)$,
corresponds to the Fourier component of the vector potential generated by an
orbiting charge in a homogeneous medium with dielectric permittivity $%
\varepsilon _{j}$. The coefficients $C_{(j)n}^{(p)}$ are expressed in terms
of $C_{n}^{(2p)}$, given above, as%
\begin{equation}
C_{(j)n}^{(p)}=\frac{C_{n}^{(2p)}}{V_{n+p}^{JH}}|_{r^{\prime }=r_{0},\omega
=\omega _{n}},  \label{C2pie}
\end{equation}%
for the region with dielectric permittivity $\varepsilon _{j}$. Here, $%
C_{n}^{(2p)}$ in the exterior and interior regions are given by the first
lines of (\ref{Cni}) and (\ref{Cnme}). The explicit expressions read%
\begin{equation}
C_{(j)n}^{(p)}=\frac{H_{n+p}(\lambda _{1}r_{0})}{V_{n+p}^{JH}}\left\{ 
\begin{array}{c}
1 \\ 
\frac{i\pi }{2}V_{n+p}^{JJ}%
\end{array}%
\right\} +\frac{p\lambda _{01}J_{n}(\lambda _{01})}{2\alpha _{n}\left(
k\right) }\sum_{l=\pm 1}\frac{H_{n+l}(\lambda _{1}r_{0})}{%
V_{n+p}^{JH}V_{n+l}^{JH}}\left\{ 
\begin{array}{c}
H_{n+p}(\lambda _{11}) \\ 
J_{n+p}(\lambda _{01})%
\end{array}%
\right\} ,  \label{Cpjn}
\end{equation}%
where the first and second lines correspond to $j=0$ and $j=1$,
respectively, and $\lambda _{j}$ is defined by (\ref{lamj2}). The modes with 
$n=0$ are static and do not contribute to the radiation fields. In what
follows, the presentation will be given for $n\neq 0$.

The coefficients (\ref{Cpjn}) are connected by the relation 
\begin{equation}
C_{(1)n}^{(p)}=C_{(0)n}^{(p)}\frac{J_{n+p}(\lambda _{01})}{H_{n+p}(\lambda
_{11})}+\frac{i\pi }{2}J_{n+p}(\lambda _{11})\frac{H_{n+p}(\lambda _{10})}{%
H_{n+p}(\lambda _{11})}.  \label{relCpj}
\end{equation}%
In deriving this formula, we used the relation 
\begin{equation}
V_{n+p}^{JJ}H_{n+p}(\lambda _{11})-V_{n+p}^{JH}J_{n+p}(\lambda _{11})=-\frac{%
2i}{\pi }J_{n+p}(\lambda _{01}),  \label{relVjj}
\end{equation}%
and the Wronskian relation between the Bessel and Hankel functions. Using
the formula (\ref{relCpj}), it can be shown that the vector potential is
continuous at the separation boundary, $%
A_{m,n}^{(1)}(k,r_{1}+0)=A_{m,n}^{(0)}(k,r_{1}-0)$.

The Fourier components of the scalar potential inside and outside the
cylinder are found from the relation 
\begin{equation}
\varphi _{n}^{(j)}(k,r)=\frac{c}{\omega \varepsilon _{j}}\left\{ \frac{n}{r}%
A_{2,n}^{(j)}(k,r)-\frac{i}{r}\partial _{1}\left[ rA_{1,n}^{(j)}(k,r)\right]
\right\} .  \label{phinj}
\end{equation}%
By taking into account that%
\begin{equation}
\frac{1}{r}\partial _{r}\left[ rZ_{n+p}^{(j)}(\lambda r)\right] =p\lambda
Z_{n}^{(j)}(\lambda r)-\frac{pn}{r}Z_{n+p}^{(j)}(\lambda r),  \label{Zjrel}
\end{equation}%
for the functions (\ref{Znu}), we get%
\begin{equation}
\varphi _{n}^{(j)}(k,r)=\varphi _{n}^{(\varepsilon _{1})}(k,r)\delta _{1j}-%
\frac{qv\lambda _{j}}{2\pi \omega _{n}\varepsilon _{j}}%
\sum_{p}C_{(j)n}^{(p)}Z_{n}^{(j)}(\lambda _{j}r).  \label{phinj2}
\end{equation}%
Here, 
\begin{equation}
\varphi _{n}^{(\varepsilon _{j})}(k,r)=\frac{iq}{2\varepsilon _{j}}%
J_{n}(\lambda _{j}r_{0<})H_{n}(\lambda _{j}r_{0>})  \label{phinhom}
\end{equation}%
corresponds to the scalar potential in a homogeneous medium with
permittivity $\varepsilon _{j}$. Using the relation (\ref{relCpj}) and 
\begin{equation}
V_{n+p}^{JH}J_{n}(\lambda _{11})-V_{n+p}^{JJ}H_{n}(\lambda _{11})=\frac{%
2i\lambda _{01}}{\pi \lambda _{11}}J_{n}(\lambda _{01}),  \label{relVjj2}
\end{equation}%
we verify the continuity of the scalar potential at $r=r_{1}$: $\varphi
_{n}^{(1)}(k,r_{1}+0)=\varphi _{n}^{(0)}(k,r_{1}-0)$.

\subsection{Electric and magnetic fields}

Having the potentials, we turn to the magnetic and electric fields. The
magnetic field is found from the relation $\mathbf{B}=\mathbf{\nabla }\times 
\mathbf{A}$. By using the expressions for the Fourier component of the
vector potential, the corresponding formulas for the magnetic field read%
\begin{align}
B_{m,n}^{(j)}(k,r) &=B_{m,n}^{(\varepsilon _{1})}(k,r)\delta _{1j}+\frac{iqvk%
}{2\pi c}\sum_{p}\frac{C_{(j)n}^{(p)}}{\left( ip\right) ^{m-1}}%
Z_{n+p}^{(j)}(\lambda _{j}r),\;m=1,2,  \notag \\
B_{3,n}^{(j)}(k,r) &=B_{3,n}^{(\varepsilon _{1})}(k,r)\delta _{1j}-\frac{%
qv\lambda _{j}}{2\pi c}\sum_{p}pC_{(j)n}^{(p)}Z_{n}^{(j)}(\lambda _{j}r),
\label{Hc}
\end{align}%
with $j=0,1$ and $m=1,2$. The parts in (\ref{Hc}) corresponding to the
magnetic fields in a homogeneous medium with dielectric permittivity $%
\varepsilon _{j}$ are given by the expressions 
\begin{align}
B_{m,n}^{(\varepsilon _{j})}(k,r) &=\frac{qvk}{4c}\sum_{p}\frac{%
H_{n+p}(\lambda _{j}r_{0>})}{\left( ip\right) ^{m-1}}J_{n+p}(\lambda
_{j}r_{0<}),  \notag \\
B_{3,n}^{(\varepsilon _{j})}(k,r) &=\frac{qv\lambda _{j}}{2ic}\left\{ 
\begin{array}{cc}
J_{n}^{\prime }(\lambda _{j}r_{0})H_{n}(\lambda _{j}r), & r>r_{0} \\ 
H_{n}^{\prime }(\lambda _{j}r_{0})J_{n}(\lambda _{j}r), & r<r_{0}%
\end{array}%
\right. ,  \label{B3Hom}
\end{align}%
with $m=1,2$. Based on the relations (\ref{relCpj}), (\ref{relVjj}), and (%
\ref{relVjj2}), we can verify that, as expected, the magnetic field is
continuous on the boundary, $%
B_{m,n}^{(1)}(k,r_{1}+0)=B_{m,n}^{(0)}(k,r_{1}-0)$ for $m=1,2,3$.

The electric field strength is found using the expressions for the
potentials, or Maxwell's equation $\omega \varepsilon _{j}\mathbf{E}(\omega ,%
\mathbf{r})/c=i\mathbf{\nabla }\times \mathbf{H}(\omega ,\mathbf{r})$ for
the spectral components. For the Fourier components, defined in accordance
with (\ref{Xexpc}), this gives%
\begin{align}
E_{m,n}^{(j)}(k,r) &=E_{m,n}^{(\varepsilon _{1})}(k,r)\delta _{1j}+\frac{iqv%
}{4\pi \omega _{n}\varepsilon _{j}}\sum_{p,l=\pm 1}\left( pl\frac{\omega
_{n}^{2}}{c^{2}}\varepsilon _{j}+k^{2}\right) \frac{Z_{n+p}^{(j)}(\lambda
_{j}r)}{\left( ip\right) ^{m}}C_{(j)n}^{(l)},  \notag \\
E_{3,n}^{(j)}(k,r) &=E_{3,n}^{(\varepsilon _{1})}(k,r)\delta _{1j}+\frac{%
iqv\lambda _{j}k}{2\pi \omega _{n}\varepsilon _{j}}%
\sum_{p}C_{(j)n}^{(p)}Z_{n}^{(j)}(\lambda _{j}r).  \label{Ec}
\end{align}
The components of the electric field in a homogeneous medium of permittivity 
$\varepsilon _{j}$ are expressed as 
\begin{align}
E_{m,n}^{(\varepsilon _{j})}(k,r) &=\frac{qv}{8\omega _{n}\varepsilon _{j}}%
\sum_{p,l=\pm 1}\left( pl\frac{\omega _{n}^{2}}{c^{2}}\varepsilon
_{j}+k^{2}\right) \frac{J_{n+p}(\lambda _{j}r_{0<})}{\left( ip\right) ^{m}}%
H_{n+l}(\lambda _{j}r_{0>}),  \notag \\
E_{3,n}^{(\varepsilon _{j})}(k,r) &=\frac{qk}{2\varepsilon _{j}}%
J_{n}(\lambda _{j}r_{0<})H_{n}(\lambda _{j}r_{0>}).  \label{E3Hom}
\end{align}%
As before, in (\ref{Ec}) and (\ref{E3Hom}), $j=0,1$ and $m=1,2$. The
expressions (\ref{Hc}) and (\ref{Ec}) are valid for general case of the
dielectric permittivities dispersion. Inside the cylinder we have $j=0$ and
the axial components of the electric and magnetic fields vanish on the
cylinder axis, $B_{3,n}^{(0)}(k,0)=E_{3,n}^{(0)}(k,0)=0$, $n\neq 0$.

By using the relations (\ref{relCpj}), (\ref{relVjj}), (\ref{relVjj2}), and
the definition (\ref{alfn}) of the function $\alpha _{n}\left( k\right) $,
after lengthy but straightforward calculations, it can be shown that the
axial component of the electric field is continuous at $r=r_{1}$: $%
E_{3,n}^{(1)}(k,r_{1}+0)=E_{3,n}^{(0)}(k,r_{1}-0)$. To see the boundary
conditions on the surface of the cylinder for the azimuthal component, it is
convenient to use the relation $\mathbf{E}(\omega ,r)=i\omega \mathbf{A}%
(\omega ,r)/c-\mathbf{\nabla }\varphi (\omega ,r)$ between the spectral
components of the electric field and potentials. From here it follows that $%
E_{2,n}^{(j)}(k,r)=i\omega _{n}A_{2,n}^{(j)}(k,r)/c-in\varphi
_{n}^{(j)}(k,r)/r$. By taking into account that the fields $%
A_{2,n}^{(j)}(k,r)$ and $\varphi _{n}^{(j)}(k,r)$ are continuous on the
boundary, we conclude that the same holds for the azimuthal component of the
electric field: $E_{2,n}^{(1)}(k,r_{1}+0)=E_{2,n}^{(0)}(k,r_{1}-0)$. The
boundary condition for the radial component of the electric field is most
easily obtained by using the relation $\omega \varepsilon
_{j}E_{1,n}^{(j)}(k,r)/c=kB_{2,n}^{(j)}(k,r)-nB_{3,n}^{(j)}(k,r)(\omega ,%
\mathbf{r})/r$. The magnetic field is continuous on the boundary and we get
the standard boundary condition for the normal component of the electric
field: $\varepsilon _{1}E_{1,n}^{(1)}(k,r_{1}+0)=\varepsilon
_{0}E_{1,n}^{(0)}(k,r_{1}-0)$.

\section{Radiation at large distances from the cylinder}

\label{sec: RadInf}

First we consider the radiation at large distances from the cylinder. As
discussed above, it is determined by the modes for which $\omega
^{2}\varepsilon _{1}>c^{2}k^{2}$. The radial dependence of these modes is
described by the Hankel function $H_{n}(\lambda _{1}r)$ with $\lambda _{1}>0$
given by (\ref{lamj2}). At large distances from the cylinder, $\lambda
_{1}r\gg 1$, one has $H_{n}(y)\approx \sqrt{2/\pi y}e^{iy-n\pi /2-\pi /4}$,
with $y=\lambda _{1}r$, and the spacetime dependence of the Fourier
components of the fields is given by $e^{i\lambda _{1}r+ikz+in\left( \phi
-\omega _{0}t\right) }/\sqrt{r}$. These components describe radiation fields
with angular frequency $\omega _{n}=|n|\omega _{0}$ propagating along the
direction forming angle $\theta $ with respect to the axis $z$, determined
from 
\begin{equation}
k=\frac{\omega _{n}}{c}\sqrt{\varepsilon _{1}}\cos \theta .  \label{kcos}
\end{equation}%
The energy flux per unit time and through the cylindrical surface of radius $%
r$, averaged over the rotation period $T=2\pi /\omega _{0}$, is given by 
\begin{equation}
P=\frac{c}{2T}\int_{0}^{T}dt\int_{-\infty }^{\infty }dz\,r\,\mathbf{n}%
_{r}\cdot \left[ \mathbf{E}\times \mathbf{B}\right] ,  \label{FluxI}
\end{equation}%
where $\mathbf{n}_{r}$ is the exterior unit vector normal to the integration
surface. Substituting the Fourier expansions (\ref{Xexpc}) we find%
\begin{equation}
P=\pi rc\sum_{n=0}^{\infty }\left( 2-\delta _{0n}\right) \int_{-\infty
}^{+\infty }dk\,\,\mathbf{n}_{r}\cdot \mathrm{Re}\,\left[ \mathbf{E}%
_{n}(k,r)\times \,\mathbf{B}_{n}^{\ast }(k,r)\right] .  \label{I1}
\end{equation}%
Taking the limit $r\rightarrow \infty $, one obtains%
\begin{equation}
P=\frac{1}{4}q^{2}v^{2}\sum_{n=1}^{\infty }\int \frac{dk}{\varepsilon
_{1}\omega _{n}}\left[ \frac{\omega _{n}^{2}}{c^{2}}\varepsilon
_{1}\left\vert W_{n}^{(+1)}-W_{n}^{(-1)}\right\vert ^{2}+k^{2}\left\vert
W_{n}^{(+1)}+W_{n}^{(-1)}\right\vert ^{2}\right] ,  \label{I2}
\end{equation}%
where the integration over $k$ goes within the region $-\omega _{n}\sqrt{%
\varepsilon _{1}}/c<k<\omega _{n}\sqrt{\varepsilon _{1}}/c$ and the notation 
$W_{n}^{(p)}=J_{n+p}(\lambda _{1}r_{0})+2iC_{(j)n}^{(p)}/\pi $ is
introduced. The explicit expression reads%
\begin{equation}
W_{n}^{(p)}=J_{n+p}(\lambda _{1}r_{0})-\frac{V_{n+p}^{JJ}}{V_{n+p}^{JH}}%
H_{n+p}(\lambda _{1}r_{0})+\frac{ip\lambda _{01}J_{n}(\lambda _{01})}{\pi
\alpha _{n}\left( k\right) }\frac{J_{n+p}(\lambda _{01})}{V_{n+p}^{JH}}%
\sum_{l=\pm 1}\frac{H_{n+l}(\lambda _{1}r_{0})}{V_{n+l}^{JH}}.  \label{Wpn}
\end{equation}%
Note that, from (\ref{VFG2}) with $i=1$ and $F=J$, for the combinations of
the Bessel and Hankel functions in (\ref{Wpn}) we have 
\begin{equation}
V_{n+p}^{JG}=p\left[ \lambda _{11}J_{n+p}(\lambda _{01})G_{n}(\lambda
_{11})-\lambda _{01}J_{n}(\lambda _{01})G_{n+p}(\lambda _{11})\right] ,
\label{VJG}
\end{equation}%
with $G=J,H$.

In (\ref{I2}), we can pass to the integration over the angle $\theta $, $%
-\pi <\theta <\pi $, in accordance with (\ref{kcos}). Introducing the
average power radiated by the charge at a harmonic $n$ into a unit solid
angle, $dP_{n}/d\Omega $, in accordance with $P=\sum_{n=1}^{\infty }\int
d\Omega \,(dP_{n}/d\Omega )$, where $d\Omega =\sin \theta d\theta d\phi $ is
the solid angle element, we get (see also \cite{Grig95,Kota00})%
\begin{equation}
\frac{dP_{n}}{d\Omega }=\frac{q^{2}\beta ^{2}\omega _{n}^{2}\sqrt{%
\varepsilon _{1}}}{8\pi c}\left[ \left\vert
W_{n}^{(1)}-W_{n}^{(-1)}\right\vert ^{2}+\left\vert
W_{n}^{(1)}+W_{n}^{(-1)}\right\vert ^{2}\cos ^{2}\theta \right] ,  \label{I3}
\end{equation}%
with $\beta =v/c$. The quantities in the arguments of the cylinder functions
in (\ref{Wpn}) are expressed as%
\begin{equation}
\lambda _{0}=\frac{\omega _{n}}{c}\sqrt{\varepsilon _{0}-\varepsilon
_{1}\cos ^{2}\theta },\;\lambda _{1}=\frac{\omega _{n}}{c}\sqrt{\varepsilon
_{1}}\sin \theta .  \label{lamb01}
\end{equation}%
The formula (\ref{I3}) is valid for general case of complex dielectric
function $\varepsilon _{0}=\varepsilon _{0}(\omega )$. Note that, for a
given angular velocity $\omega _{0}$, the dependence on the radius of the
rotation orbit enters in the formula (\ref{I3}) via the Hankel functions $%
H_{n\pm 1}(\lambda _{1}r_{0})$. The spectral-angular density of the
radiation for a charge circulating inside a cylinder has been studied in 
\cite{Kota02b}.

For $\varepsilon _{0}=\varepsilon _{1}$ one has $W_{n}^{(p)}=J_{n+p}(\lambda
_{1}r_{0})$ and from (\ref{I3}) we obtain the corresponding quantity for a
charge circulating in a homogeneous medium with permittivity $\varepsilon
_{1}$ (see, e.g., \cite{Zrel70}):%
\begin{equation}
\frac{dP_{n}^{(\varepsilon _{1})}}{d\Omega }=\frac{q^{2}\omega _{n}^{2}}{%
2\pi c\sqrt{\varepsilon _{1}}}\,\left[ \beta ^{2}\varepsilon
_{1}J_{n}^{\prime 2}(n\beta \sqrt{\varepsilon _{1}}\sin \theta )+\cot
^{2}\theta J_{n}^{2}(n\beta \sqrt{\varepsilon _{1}}\sin \theta )\right] .
\label{Ihom}
\end{equation}%
Note that $\omega _{n}r_{0}=nv$ and the spectral-angular density of the
radiated power (\ref{I3}) has a functional structure 
\begin{equation}
\frac{dP_{n}}{d\Omega }=\frac{f_{n}(\beta _{1},r_{1}/r_{0},\varepsilon
_{0}/\varepsilon _{1})}{r_{0}^{2}\sqrt{\varepsilon _{1}}},\;\beta _{1}=\frac{%
v}{c}\sqrt{\varepsilon _{1}}  \label{Pstruc}
\end{equation}%
This means that for fixed $\beta _{1}$, the combination $\sqrt{\varepsilon
_{1}}r_{0}^{2}dP_{n}/d\Omega $ depends on $r_{1}$ and $r_{0}$ and on the
dielectric permittivities in the form of the ratios $r_{1}/r_{0}$ and $%
\varepsilon _{0}/\varepsilon _{1}$. In the limit $r_{1}/r_{0}\ll 1$, to the
leading order, we get $dP_{n}/d\Omega \approx dP_{n}^{(\varepsilon
_{1})}/d\Omega $ and the contribution induced by the presence of the
cylinder decays as $(r_{1}/r_{0})^{2n}$. For a non-relativistic motion, $%
\beta \ll 1$, one has $dP_{n}/d\Omega \varpropto \beta ^{2(n+1)}$ and the
dominant contribution comes from the lowest harmonic $n=1$. The radiation
features for a charged particle circulating in a homogeneous medium are
discussed in \cite{Zrel70},\cite{Kita60}-\cite{Boni86}, based on the formula
(\ref{Ihom}). It has been shown that, under the Cherenkov condition $\beta
_{1}>1$, the interference of synchrotron and Cherenkov radiations leads to
interesting effects in both spectral and angular distributions of the
radiation intensity.

According to the analysis in \ref{sec:Features}, narrow peaks appear when $%
\lambda _{11}<n<\lambda _{01}$. However, this is not sufficient. Other
factors in (\ref{Wpn}) need to be estimated. As mentioned above, at the
zeros of the function (\ref{alfan0}) one has $\alpha _{n}\left( k\right)
\varpropto e^{-2n\zeta (\lambda _{11}/n)}$. Under the condition $\lambda
_{01}>n$, which is necessary for the zeros of $\alpha _{n}^{(0)}\left(
k\right) $, one has $V_{n+p}^{JH}\varpropto e^{n\zeta (\lambda _{11}/n)}$.
From (\ref{Wpn}) it follows that, for the peaks to appear, one must have $%
\lambda _{1}r_{0}<n$, for which $H_{n+l}(\lambda _{1}r_{0})\varpropto
e^{n\zeta (\lambda _{1}r_{0}/n)}$. In this case, at the peaks $%
dP_{n}/d\Omega \varpropto e^{2n\zeta (\lambda _{1}r_{0}/n)}$. Consequently,
the conditions for the appearance of the peaks read $\lambda
_{1}r_{0}<n<\lambda _{01}$. With the expressions (\ref{lamb01}), they are
translated to%
\begin{equation}
1-\frac{c^{2}}{v^{2}\varepsilon _{1}}<\cos ^{2}\theta <\frac{\varepsilon _{0}%
}{\varepsilon _{1}}\left( 1-\frac{c^{2}}{v_{c}^{2}\varepsilon _{0}}\right) ,
\label{peakscond3}
\end{equation}%
where $v_{c}=r_{1}\omega _{0}=vr_{1}/r_{0}$ is the velocity of the particle
image on the cylinder surface. In particular, the conditions $\varepsilon
_{0}>\varepsilon _{1}$ and $v_{c}\sqrt{\varepsilon _{0}}>c$ should be
obeyed. The latter relation states that the velocity of the charge image on
the cylinder surface should obey the Cherenkov condition for the material of
the cylinder.

Using the asymptotic expressions for cylinder functions allows us to
estimate the widths of the peaks. By expanding the function $\alpha
_{n}\left( k\right) $ in (\ref{Wpn}) near the peaks, it can be seen that the
width of the peak is determined by the imaginary part of the function $%
\alpha _{n}\left( k\right) $, evaluated at the location of the peak. The
width is of the order $e^{-2n\zeta (\lambda _{11}/n)}$. Note that the
function $\zeta (y)$ is monotonically decreasing with increasing $y$ and $%
\zeta (\lambda _{11}/n)>\zeta (\lambda _{1}r_{0}/n)$. We have considered an
idealized case in which the dielectric permittivities are real. In real
situations, the increase in the peak height with increasing $n$ is
restricted by several factors. In particular, the imaginary parts of the
permittivites contribute to the expansion of the function $\alpha _{n}\left(
k\right) $ near the peaks, restricting the increase in height and decrease
in width.

\begin{figure}[tbph]
\begin{center}
\begin{tabular}{cc}
\epsfig{figure=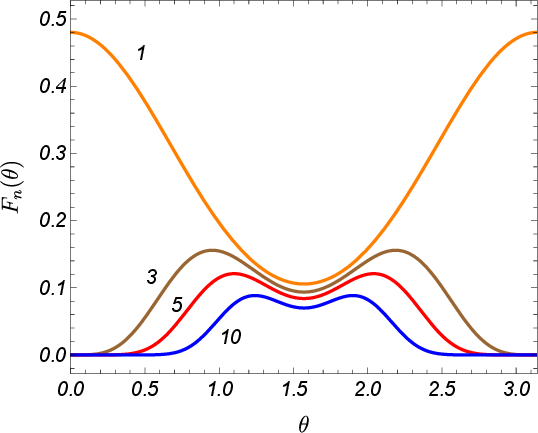,width=8cm,height=6.5cm} & \quad %
\epsfig{figure=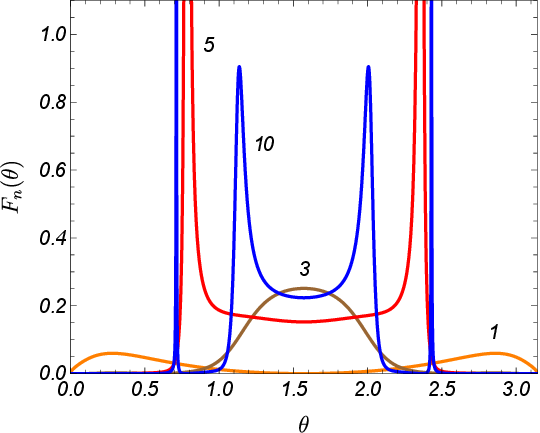,width=8cm,height=6.5cm}%
\end{tabular}%
\end{center}
\caption{The angular density of the number of the radiated quanta per
rotation period, in units of $q^{2}/\hbar c$, versus the angle $\protect%
\theta $, for different values of the radiation harmonic $n$ (numbers near
the curves). The left panel corresponds to the synchrotron radiation in a
vacuum ($\protect\varepsilon _{0}=\protect\varepsilon _{1}=1$) and the right
panel presents the data for a charge rotating around a dielectric cylinder
with permittivity $\protect\varepsilon _{0}=3.8$ ($\protect\varepsilon %
_{1}=1 $). The graphs are plotted for $\protect\beta =0.98$, $%
r_{1}/r_{0}=0.95$.}
\label{fig1}
\end{figure}

In the numerical examples we will consider the angular dependence of the the
number of the radiated quanta per unit time and on a given harmonic,
averaged over the rotation period. It is given by $dN_{n}/d\Omega =(\hbar
\omega _{n})^{-1}dP_{n}/d\Omega $. In Fig. \ref{fig1}, we display the
angular distribution of the number of the radiated quanta (in units of $%
q^{2}/\hbar c$) during the rotation period $T=2\pi /\omega _{0}$, given by%
\begin{equation}
F_{n}(\theta )=T\frac{\hbar c}{q^{2}}\frac{dN_{n}}{d\Omega }.  \label{Fntet}
\end{equation}%
The right panel corresponds to the radiation from a charge rotating in a
vacuum ($\varepsilon _{1}=1$) around a cylinder with dielectric permittivity 
$\varepsilon _{0}=3.8$ (real part of the dielectric permiitivity for fused
silica). For the values of the other parameters we have taken $\beta =0.98$, 
$r_{1}/r_{0}=0.95$, and the numbers near the curves correspond to the values
of the harmonic $n$. The angular distribution is symmetric with respect to
the plane $\theta =\pi /2$. For comparison, the left panel of Fig. \ref{fig1}
shows the corresponding quantity for the synchrotron radiation from a charge
circulating in a vacuum ($\varepsilon _{0}=\varepsilon _{1}=1$). We have
verified that the locations of the peaks in the problem with a cylinder are
given with high accuracy by zeros of the function (\ref{alfan0}). For
example, the peak for the harmonic $n=5$ with the height $F_{n}(\theta
)\approx 8.4$ is located at $\theta \approx 0.78$ and the same approximate
value is predicted by the equation (\ref{PeakEq}). One has the same accuracy
for the case $n=10$ with the peaks at $\theta \approx 0.71$ and $1.13$. For
the height of the peak at $\theta \approx 0.71$ one has $F_{n}(\theta
)\approx 1.5$.

In the example of Fig. \ref{fig1}, the charge moves in a vacuum and this
figure illustrates the influence of the dielectric cylinder on the
characteristics of the synchrotron radiation. The right panel of Fig. \ref%
{fig2} presents the case with a charge circulating in a medium with
dielectric permittivity $\varepsilon _{1}=2.1$ (real part of dielectric
permittivity for Teflon). The dielectric permittivity for the cylinder and
the values of the other parameters are the same as those for Fig. \ref{fig1}%
. The graphs on the left panel correspond to the radiation in a homogeneous
medium with permittivity $\varepsilon _{1}$. Now, the Cherenkov condition in
the exterior medium is satisfied, $\beta _{1}>1$, and the radiation outside
the cylinder is a superposition of synchrotron and Cherenkov radiations.
Again, we have checked that the positions of the peaks on the harmonics $%
n=5,10$ are determined by the roots of the equation (\ref{PeakEq}). The
height of the peak for $n=5$ located at $\theta \approx 0.38$ is equal to $%
\approx 2.54$. For the radiation on the harmonic $n=10$, the heights of the
peaks at $\theta \approx 0.25$ and $\theta \approx 0.56$ are given by $%
F_{n}(\theta )\approx 4.8\cdot 10^{4}$ and $F_{n}(\theta )\approx 29.8$,
respectively. Of course, these data are for an idealized problem with real
dielectric permittivities. As mentioned above, the imaginary parts of the
permittivities will reduce the heights of the peaks. In the case of a
particle rotating in a homogeneous medium ($\varepsilon _{0}=\varepsilon
_{1} $), the only nonzero contribution to the angular density of the
radiation intensity in the limit $\theta \rightarrow 0$ comes from the mode $%
n=1$. For a particle circulating around a cylinder ($\varepsilon _{0}\neq
\varepsilon _{1}$), we have $\lim_{\theta \rightarrow 0}dP_{n}/d\Omega =0$
for all harmonics. This behavior is seen from the expression (\ref{Wpn}) in
the limit $\lambda _{1}\rightarrow 0$, corresponding to $\theta \rightarrow
0 $. For a charge rotating in a homogeneous medium, the function $%
W_{n}^{(p)} $ is reduced to $J_{n+p}(\lambda _{1}r_{0})$ and the mode $n=1$
with $p=-1$ gives a nonzero contribution in the limit at hand. In the
problem with a dielectric cylinder, the last term in (\ref{Wpn}) tends to
zero in the limit $\lambda _{1}\rightarrow 0$ for all modes, and the
contribution of the first term with the function $J_{n+p}(\lambda _{1}r_{0})$
is canceled by the second term.

\begin{figure}[tbph]
\begin{center}
\begin{tabular}{cc}
\epsfig{figure=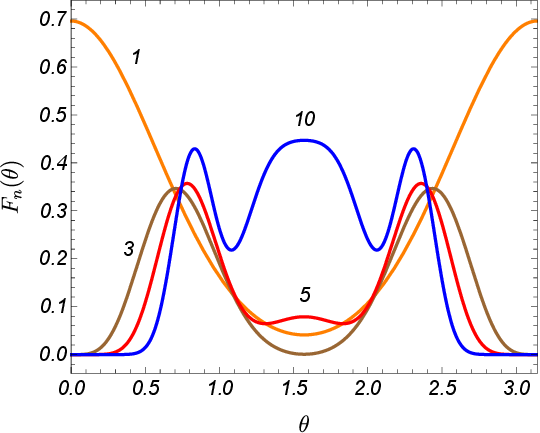,width=8cm,height=6.5cm} & \quad %
\epsfig{figure=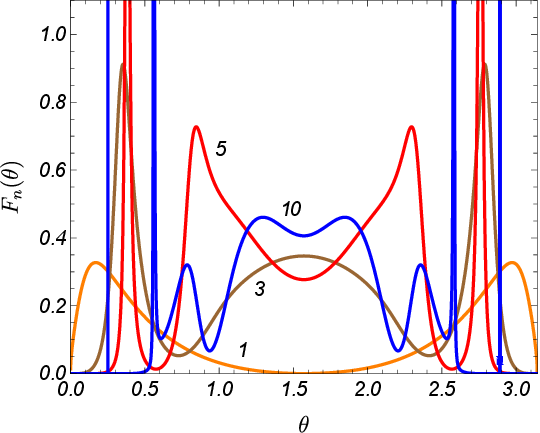,width=8cm,height=6.5cm}%
\end{tabular}%
\end{center}
\caption{The same as in Fig. \protect\ref{fig1} for $\protect\varepsilon %
_{1}=2.1$.}
\label{fig2}
\end{figure}

\begin{figure}[tbph]
\begin{center}
\epsfig{figure=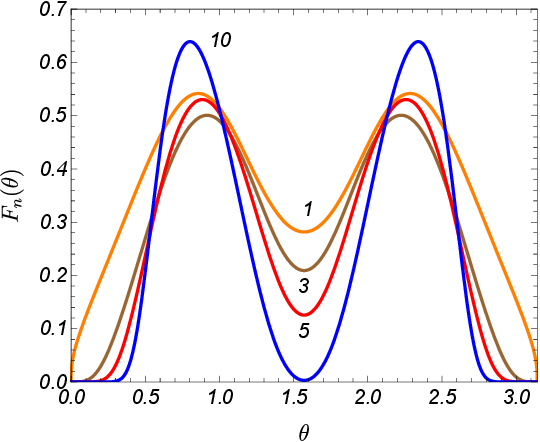,width=8cm,height=6.5cm}
\end{center}
\caption{The same as on the right panel of Fig. \protect\ref{fig1} for $%
\protect\varepsilon _{0}=1$, $\protect\varepsilon _{1}=3.8$. }
\label{fig3}
\end{figure}

The discussion in Section \ref{sec:Features} showed that the condition $%
\varepsilon _{0}>\varepsilon _{1}$ is necessary for peaks to appear. In Fig. %
\ref{fig3}, the function $F_{n}(\theta )$ is plotted for different values of 
$n$ (numbers near the curves) and for $\varepsilon _{0}=1$, $\varepsilon
_{1}=3.8$ (cylindrical hole in a dielectric medium). For the other
parameters we have taken the same values as in Figs. \ref{fig1} and \ref%
{fig2}. According to the analysis presented above, in this case the strong
peaks in the angular distribution are absent.

We have considered numerical examples with a positive dielectric function of
the cylinder. Formula (\ref{I3}) for the angular density of the radiation
intensity is also valid in spectral ranges where the permittivity $%
\varepsilon _{0}$ becomes negative. The arguments given above for the
presence of strong peaks in the angular distribution of the radiation
intensity also apply in this case. The corresponding features of the
spectral-angular distribution of the radiation at large distances from the
cylinder have been discussed in \cite{Saha20b}.

\section{Radiation on the normal modes of a cylinder}

\label{sec:RadGuid}

\subsection{Radiation fields}

Now, let us consider the radiation in the spectral range where $\lambda
_{1}^{2}<0$ and the radiation is confined inside the cylinder or near its
surface. In order to separate the corresponding contributions to the total
field, we consider the limit $z\rightarrow \infty $ in the expansion (\ref%
{Xexpc}). The problem under consideration is symmetric with respect to the
plane $z=0$ and, without loss of generality, we can consider the half space $%
z>0$. The phase of the exponent in the integrand of (\ref{Xexpc}) has no
stationary points. According to the stationary phase method, for regular
functions $X_{n}(k,r)$ and for large $z$, the integral would decay
exponentially. The radiation fields will come from possible singular points
of the integrand. For the fields discussed in Section \ref{sec:Circ}, the
singularities are contained in the parts with the function $\alpha
_{n}\left( k\right) $ and they are poles at the zeros of this function. As
mentioned above, they present the eigenmodes of a cylindrical waveguide.

Let $\pm k_{n,s}$, $k_{n,s}>0$, $s=1,2,...$, be the solutions of the
equation (\ref{alfn0}) with respect to $k$ for $n\neq 0$. For $\lambda
_{11}^{2}<0$ the equation is transformed to the form (\ref{NormMode2}). For
the roots $k_{n,s}$ we introduce the notations 
\begin{equation}
\lambda _{n,s}^{(0)}\equiv r_{1}\sqrt{\omega _{n}^{2}\varepsilon
_{0}/c^{2}-k_{n,s}^{2}},\;\gamma _{n,s}^{(1)}\equiv r_{1}\sqrt{%
k_{n,s}^{2}-\varepsilon _{1}\omega _{n}^{2}/c^{2}}.  \label{lams}
\end{equation}%
For real $\varepsilon _{j}$, $\gamma _{n,s}^{(1)}$ is positive, while $%
\lambda _{n,s}^{(0)}$ can be either positive or purely imaginary. For the
evaluation of the radiation parts in the fields, we need to specify the
integration contour in (\ref{Xexpc}) near the poles $\pm k_{n,s}$. This is
done introducing a small imaginary part of the dielectric permittivity, $%
\varepsilon _{j}=\varepsilon _{j}^{\prime }+i\varepsilon _{j}^{\prime \prime
}$, with $\varepsilon _{j}^{\prime \prime }(\omega )>0$ for $\omega >0$. On
the base of this, it can be seen that (for details see, e.g., \cite{Saha12})
in the integral over $k$ the contour should avoid the poles $k_{z}=k_{n,s}$ (%
$k_{z}=-k_{n,s}$) from below (above). With this specification and for $z>0$,
closing the integration contour by the large semicircle in the upper
half-plane of complex variable $k$, the radiation fields are obtained with
the help of the residue theorem: 
\begin{equation}
X^{\mathrm{(r)}}(x)=4\pi {\mathrm{Re}}\left\{ i\sum_{n=1}^{\infty
}e^{in(\phi -\omega _{0}t)}\sum_{s}\underset{k=k_{n,s}}{\mathrm{Res}}\,\left[
e^{ikz}X_{n}(k,r)\right] \right\} ,  \label{Xrad}
\end{equation}%
where the superscript (r) stands for the radiation parts of the fields. Note
that the nonzero contributions to the residue come from the parts of the
coefficients $C_{(j)n}^{(p)}$ containing the factor $1/\alpha _{n}\left(
k\right) $.

Evaluating the residue, the radiation fields are presented in the form 
\begin{equation}
X^{\mathrm{(r)}}(x)=\frac{qv}{c}\sum_{n=1}^{\infty }\sum_{s=1}^{s_{n}}\frac{%
X_{n,s}(r)}{\alpha _{n}^{\prime }\left( k_{n,s}\right) }R\left[ n(\phi
-\omega _{0}t)+k_{n,s}z\right] ,  \label{Xrad2}
\end{equation}%
where $s_{n}$ is the number of the modes with given $n$, and 
\begin{equation}
\alpha _{n}^{\prime }\left( k_{n,s}\right) =\partial _{k}\alpha _{n}\left(
k\right) |_{k=k_{n,s}}.  \label{alfder}
\end{equation}%
For the function determining the dependence on the spacetime coordinates $%
(t,\phi ,z)$ one has $R(x)=\sin x$ for the components $X^{\mathrm{(r)}%
}=E_{1}^{\mathrm{(r)}},B_{2}^{\mathrm{(r)}},B_{3}^{\mathrm{(r)}}$ and $%
R(x)=\cos x$ for the components $X^{\mathrm{(r)}}=E_{2}^{\mathrm{(r)}%
},E_{3}^{\mathrm{(r)}},B_{1}^{\mathrm{(r)}}$.

The parts in the coefficients $C_{(j)n}^{(p)}$, giving the contributions to
the radiation fields are expressed in terms of the function%
\begin{equation}
U_{(j)n}^{(p)}=\frac{\lambda _{01}J_{n}(\lambda _{01})}{V_{n+p}^{JK}}%
\sum_{l=\pm 1}\frac{K_{n+l}(\gamma _{1}r_{0})}{V_{n+l}^{JK}}\left\{ 
\begin{array}{cc}
K_{n+p}(\gamma _{11}), & j=0 \\ 
J_{n+p}(\lambda _{01}), & j=1%
\end{array}%
\right. ,  \label{Upj}
\end{equation}%
with the notation%
\begin{align}
V_{n}^{JK} &=J_{n}\left( \lambda _{0}r_{1}\right) r_{1}\partial
_{r_{1}}K_{n}\left( \gamma _{1}r_{1}\right) -K_{n}\left( \gamma
_{1}r_{1}\right) r_{1}\partial _{r_{1}}J_{n}\left( \lambda _{0}r_{1}\right) 
\notag \\
&=p\lambda _{01}J_{n+p}(\lambda _{01})K_{n}\left( \gamma _{11}\right)
-\gamma _{11}J_{n}\left( \lambda _{01}\right) K_{n+p}(\gamma _{11}),
\label{Vjk}
\end{align}%
where $p=\pm 1$. Note that we have the relation%
\begin{equation}
V_{n+p}^{JK}=pJ_{n}\left( \lambda _{01}\right) K_{n}\left( \gamma
_{11}\right) \left( V_{n}-pnu\right) ,  \label{VJKVn}
\end{equation}%
with the function $V_{n}$ from (\ref{Vn}). For the modes with $\lambda
_{1}^{2}<0$, the function (\ref{alfn}) is written as%
\begin{equation}
\alpha _{n}\left( k\right) =\frac{\varepsilon _{0}}{\varepsilon
_{1}-\varepsilon _{0}}-\frac{\lambda _{01}}{2}J_{n}(\lambda
_{01})\sum_{l=\pm 1}l\frac{K_{n+l}(\lambda _{11})}{V_{n+l}^{JK}}.
\label{alfng}
\end{equation}%
By using the properties of the Bessel functions and the equation $\alpha
_{n}\left( k_{n,s}\right) =0$ for the modes, the derivative of the function (%
\ref{alfng}) at the roots $k=k_{n,s}$ is presented in the form%
\begin{align}
\alpha _{n}^{\prime }\left( k\right) &=\frac{r_{1}^{3}k}{2}\sum_{l=\pm
1}\left( \frac{J_{n}K_{n}}{V_{n+l}^{JK}}\right) ^{2}\left\{ \gamma _{1}\frac{%
K_{n+l}}{K_{n}}\left[ 1-\left( \frac{n+l}{\lambda _{01}}\right) ^{2}+\left( 
\frac{J_{n}^{\prime }}{J_{n}}+\frac{1}{\lambda _{01}}\right) ^{2}\right]
\right.  \notag \\
& \left. -l\lambda _{0}\frac{J_{n+l}}{J_{n}}\left[ 1+\left( \frac{n+l}{%
\gamma _{11}}\right) ^{2}-\left( \frac{K_{n}^{\prime }}{K_{n}}+\frac{1}{%
\gamma _{11}}\right) ^{2}\right] \right\} _{k=k_{n,s}}.  \label{alfnder}
\end{align}

In the region $r<r_{1}$, the functions $X_{n,s}(r)$ for the components of
the magnetic field read 
\begin{align}
B_{m,n,s}(r) &=-k\sum_{p}p^{m}U_{(0)n}^{(p)}J_{n+p}(\lambda
_{0}r)|_{k=k_{n,s}},  \notag \\
B_{3,n,s}(r) &=\lambda _{0}\sum_{p}U_{(0)n}^{(p)}J_{n}(\lambda
_{0}r)|_{k=k_{n,s}},  \label{H3s}
\end{align}%
with $m=1,2$. For the electric field, by using the representations (\ref{Ec}%
), we get%
\begin{align}
E_{m,n,s}(r) &=\frac{(-1)^{m}c}{2\omega _{n}\varepsilon _{0}}\sum_{p,l=\pm
1}p^{m-1}J_{n+p}(\lambda _{0}r)\left( \frac{\omega _{n}^{2}\varepsilon _{0}}{%
c^{2}}+lk^{2}\right) U_{(0)n}^{(lp)}|_{k=k_{n,s}},  \notag \\
E_{3,n,s}(r) &=-\frac{v\lambda _{0}k}{n\omega _{0}\varepsilon _{0}}%
\sum_{p}pU_{(0)n}^{(p)}J_{n}(\lambda _{0}r)|_{k=k_{n,s}}.  \label{E3s}
\end{align}%
On the axis of the waveguide we have $B_{3,n,s}(0)=E_{3,n,s}(0)=0$.

For the region outside the cylinder, $r>r_{1}$, the components of the
magnetic field are expressed as%
\begin{align}
B_{m,n,s}(r) &=-k\sum_{p}p^{m}U_{(1)n}^{(p)}K_{n+p}(\gamma
_{1}r)|_{k=k_{n,s}},  \notag \\
B_{3,n,s}(r) &=-\gamma _{1}\sum_{p}pU_{(1)n}^{(p)}K_{n}(\gamma
_{1}r)|_{k=k_{n,s}},  \label{H3se}
\end{align}%
and for the electric field we get%
\begin{align}
E_{m,n,s}(r) &=\frac{(-1)^{m}c}{2\omega _{n}\varepsilon _{1}}\sum_{p,l=\pm
1}p^{m-1}K_{n+p}(\gamma _{1}r)\left( l\frac{\omega _{n}^{2}\varepsilon _{1}}{%
c^{2}}+k^{2}\right) U_{(1)n}^{(lp)}|_{k=k_{n,s}},  \notag \\
E_{3,n,s}(r) &=\frac{c\gamma _{1}k}{\omega _{n}\varepsilon _{1}}%
\sum_{p}U_{(1)n}^{(p)}K_{n}(\gamma _{1}r)|_{k=k_{n,s}},  \label{E3se}
\end{align}%
again, with $m=1,2$. For a given angular velocity $\omega _{0}$ of the
charge rotation, the fields depend on the orbit radius $r_{0}$ through the
functions $K_{n\pm 1}(\gamma _{1}r_{0})$ in (\ref{Upj}) and they
monotonically decay with increasing $r_{0}$.

The expressions given in this section describe the fields for both guided
and surface polaritonic type modes. For guided modes, the arguments $\lambda
_{01}=\lambda _{n,s}^{(0)}$ and $\lambda _{0}r=$ $\lambda
_{n,s}^{(0)}r/r_{1} $ of the Bessel function $J_{n}(y)$ are real. For
radiated surface polaritons, these arguments are purely imaginary with a
positive imaginary part. In this case we introduce the modified Bessel
function $I_{n}(y)$.

For guided modes we have%
\begin{equation}
\varepsilon _{1}<(ck/\omega _{n})^{2}<\varepsilon _{0}.  \label{kguided}
\end{equation}%
The dispersion $\omega =\omega (k)$ for those modes is determined from the
equation (\ref{NormMode2}). For waves radiated by a charge on a given
harmonic $n$, the allowed values of $k$ are determined by the intersection
points of the dispersion curves $\omega =\omega (k)$ in the $(k,\omega )$%
-plane with the horizontal line $\omega =n\omega _{0}$. The dependence of
the radiation fields on the radius of the rotation orbit enters in the form
of the functions $K_{n\pm 1}(\gamma _{1}r_{0})$ (see (\ref{Upj})) and these
fields in both exterior and interior regions are suppressed by the factor $%
e^{-\gamma _{1}r_{0}}$ for $\gamma _{1}r_{0}\gg 1$. The radiation is a
result of medium polarization by the field of charged particle. For the
spectral range corresponding to the guided modes, the Fourier components of
the charge's field exponentially decay at large distances from the charge
trajectory and the mentioned exponential suppression of the radiation
intensity is a consequence of that behavior.

\subsection{Radiation energy flux}

Given the radiation fields, we can evaluate the radiation intensity on the
eigenmodes of a dielectric cylinder. As a physical characteristic of the
intensity we consider the energy flux per unit time through the plane $z=%
\mathrm{const}$ at large distances from the radiation source. It is
expressed in terms of the axial component of the Poynting vector and is
given by 
\begin{equation}
P^{\mathrm{(nm)}}=\frac{c}{4\pi }\int_{0}^{\infty }dr\,\int_{0}^{2\pi }d\phi
\,r\left( E_{1}^{\mathrm{(r)}}B_{2}^{\mathrm{(r)}}-E_{2}^{\mathrm{(r)}%
}B_{1}^{\mathrm{(r)}}\right) ,  \label{Ieig}
\end{equation}%
where (nm) stands for the normal modes. Plugging the radiation fields from (%
\ref{Xrad2}), with the components $X_{m,n,s}(r)$ for the electric and
magnetic fields from (\ref{H3s})-(\ref{E3se}), it can be shown that the
radial integrals are reduced to $\int_{0}^{r_{1}}dr\,rJ_{n+p}^{2}(\lambda
_{0}r)$ and $\int_{r_{1}}^{\infty }dr\,rK_{n+p}^{2}(\gamma _{1}r)$ in the
interior and exterior regions, respectively. These integrals are evaluated
by standard formulas \cite{Prud2}.

Let us denote by $P_{jn}^{\mathrm{(nm)}}$, $j=0,1$, the energy flux on a
fixed harmonic $n$ in the medium with dielectric permittivity $\varepsilon
_{j}$. After straightforward transformations and using the relation (\ref%
{VJKVn}), for the energy flux inside the cylinder one gets%
\begin{align}
P_{0n}^{\mathrm{(nm)}} &=\frac{q^{2}v^{2}r_{1}^{2}}{8\omega _{n}\varepsilon
_{0}}\sum_{s=1}^{s_{n}}\frac{k\lambda _{01}^{2}J_{n}^{2}}{\alpha
_{n}^{\prime 2}\left( k_{n,s}\right) }\left( \sum_{l^{\prime }=\pm 1}\frac{%
K_{n+l^{\prime }}\left( \gamma _{1}r_{0}\right) }{V_{n+l^{\prime }}^{JK}}%
\right) ^{2}\sum_{p,l=\pm 1}\frac{K_{n+p}}{V_{n+p}^{JK}}  \notag \\
&\times \left( \frac{\omega _{n}^{2}\varepsilon _{0}}{c^{2}}+lk^{2}\right) 
\frac{K_{n+lp}}{V_{n+lp}^{JK}}\left[ J_{n+p}^{\prime 2}+\left( 1-\left( 
\frac{n+p}{\lambda _{01}}\right) ^{2}\right) J_{n+p}^{2}\right] .
\label{I0eig}
\end{align}%
We remind that here $J_{\nu }=J_{\nu }(\lambda _{01})$ and $K_{\nu }=K_{\nu
}(\gamma _{11})$ with $\nu =n,n\pm 1$. In a similar way, the energy flux in
the exterior region, $r>r_{1}$, is presented in the form%
\begin{align}
P_{1n}^{\mathrm{(nm)}} &=\frac{q^{2}v^{2}r_{1}^{2}}{8\omega _{n}\varepsilon
_{1}}\sum_{s=1}^{s_{n}}\frac{k\lambda _{01}^{2}J_{n}^{2}}{\alpha
_{n}^{\prime 2}\left( k_{n,s}\right) }\left( \sum_{l^{\prime }=\pm 1}\frac{%
K_{n+l^{\prime }}\left( \gamma _{1}r_{0}\right) }{V_{n+l^{\prime }}^{JK}}%
\right) ^{2}\sum_{p,l=\pm 1}\frac{J_{n+p}}{V_{n+p}^{JK}}  \notag \\
&\times \left( k^{2}+l\frac{\omega _{n}^{2}\varepsilon _{1}}{c^{2}}\right) 
\frac{J_{n+lp}}{V_{n+lp}^{JK}}\left[ K_{n+p}^{\prime 2}-\left( 1+\left( 
\frac{n+p}{\gamma _{11}}\right) ^{2}\right) K_{n+p}^{2}\right] .
\label{I1eig}
\end{align}

Alternative representations are obtained by using the relations%
\begin{align}
\sum_{l=\pm 1}\left( k^{2}+l\frac{\omega _{n}^{2}\varepsilon _{1}}{c^{2}}%
\right) \frac{J_{n+lp}}{V_{n+lp}^{JK}} &=\frac{2\varepsilon _{1}\lambda _{0}%
}{r_{1}\left( \varepsilon _{0}-\varepsilon _{1}\right) }\frac{pV_{n}^{JK}}{%
K_{n}V_{n+p}^{JK}},  \notag \\
\sum_{l=\pm 1}\left( \frac{\omega _{n}^{2}\varepsilon _{0}}{c^{2}}%
+lk^{2}\right) \frac{K_{n+lp}}{V_{n+lp}^{JK}} &=\frac{2\varepsilon
_{0}\gamma _{1}}{r_{1}\left( \varepsilon _{1}-\varepsilon _{0}\right) }\frac{%
V_{n}^{JK}}{J_{n}V_{n+p}^{JK}},  \label{Rel12}
\end{align}%
where $k=k_{n,s}$. They are obtained by taking into account the equation (%
\ref{NormMode2}) with the roots $k_{n,s}$. With the relations (\ref{Rel12}),
the energy fluxes are expressed as%
\begin{align}
P_{0n}^{\mathrm{(nm)}} &=\frac{q^{2}v^{2}}{4\omega _{n}\left( \varepsilon
_{1}-\varepsilon _{0}\right) }\sum_{s=1}^{s_{n}}\frac{k\gamma _{11}\lambda
_{01}^{2}J_{n}V_{n}^{JK}}{\alpha _{n}^{\prime 2}\left( k_{n,s}\right) }\left[
\sum_{l=\pm 1}\frac{K_{n+l}\left( \gamma _{1}r_{0}\right) }{V_{n+l}^{JK}}%
\right] ^{2}  \notag \\
&\times \sum_{p=\pm 1}\frac{K_{n+p}}{\left( V_{n+p}^{JK}\right) ^{2}}\left[
J_{n+p}^{\prime 2}+\left( 1-\left( \frac{n+p}{\lambda _{01}}\right)
^{2}\right) J_{n+p}^{2}\right] _{k=k_{n,s}},  \label{I0eig2}
\end{align}%
inside the cylinder, $r<r_{1}$, and 
\begin{align}
P_{1n}^{\mathrm{(nm)}} &=\frac{q^{2}v^{2}}{4\omega _{n}\left( \varepsilon
_{0}-\varepsilon _{1}\right) }\sum_{s=1}^{s_{n}}\frac{k\lambda
_{01}^{3}J_{n}^{2}V_{n}^{JK}}{\alpha _{n}^{\prime 2}\left( k_{n,s}\right)
K_{n}}\left[ \sum_{l=\pm 1}\frac{K_{n+l}\left( \gamma _{1}r_{0}\right) }{%
V_{n+l}^{JK}}\right] ^{2}  \notag \\
&\times \sum_{p=\pm 1}\frac{pJ_{n+p}}{\left( V_{n+p}^{JK}\right) ^{2}}\left[
K_{n+p}^{\prime 2}-\left( 1+\left( \frac{n+p}{\gamma _{11}}\right)
^{2}\right) K_{n+p}^{2}\right] _{k=k_{n,s}},  \label{I1eig2}
\end{align}%
in the exterior region, $r>r_{1}$. We note that the expressions in the
square brackets in the second lines of (\ref{I0eig2}) and (\ref{I1eig2}) are
obtained from the radial integrals involving the functions $%
J_{n+p}^{2}(\lambda _{0}r)$ and $K_{n+p}^{2}(\gamma _{1}r)$, and they are
positive.

\subsection{Total energy flux and the radiated power}

Having the energy fluxes inside and outside the cylinder, the total energy
flux through the plane $z=\mathrm{const}$, for the radiation on a given
harmonic, is obtained as the sum of these two contributions:%
\begin{equation}
P_{n}^{\mathrm{(nm)}}=P_{0n}^{\mathrm{(nm)}}+P_{1n}^{\mathrm{(nm)}}.
\label{Ieigtot}
\end{equation}%
Combining the expressions (\ref{I0eig2}) and (\ref{I1eig2}) we get%
\begin{align}
P_{n}^{\mathrm{(nm)}} &=\frac{q^{2}v^{2}}{4\omega _{n}\left( \varepsilon
_{1}-\varepsilon _{0}\right) }\sum_{s=1}^{s_{n}}\frac{k\lambda
_{01}^{2}J_{n}V_{n}^{JK}}{\alpha _{n}^{\prime 2}\left( k_{n,s}\right) K_{n}}%
\left( \sum_{l=\pm 1}\frac{K_{n+l}\left( \gamma _{1}r_{0}\right) }{%
V_{n+l}^{JK}}\right) ^{2}  \notag \\
&\times \sum_{p=\pm 1}\left\{ \gamma _{11}\frac{K_{n}K_{n+p}}{\left(
V_{n+p}^{JK}\right) ^{2}}\left[ J_{n+p}^{\prime 2}+\left( 1-\left( \frac{n+p%
}{\lambda _{01}}\right) ^{2}\right) J_{n+p}^{2}\right] \right.  \notag \\
&\left. -\lambda _{01}\frac{pJ_{n}J_{n+p}}{\left( V_{n+p}^{JK}\right) ^{2}}%
\left[ K_{n+p}^{\prime 2}-\left( 1+\left( \frac{n+p}{\gamma _{11}}\right)
^{2}\right) K_{n+p}^{2}\right] \right\} _{k=k_{n,s}}.  \label{Ieig2}
\end{align}%
Using the properties of the Bessel function and by taking into account the
expression (\ref{alfnder}) for the derivative of the function $\alpha
_{n}\left( k\right) $ at the roots $k_{n,s}$, it can be shown that the sum
over $p$ is equal to $2\alpha _{n}^{\prime }(k_{n,s})/(r_{1}^{2}k_{n,s})$.
This leads to the formula%
\begin{equation}
P_{n}^{\mathrm{(nm)}}=\frac{q^{2}v^{2}}{2\omega _{n}\left( \varepsilon
_{1}-\varepsilon _{0}\right) }\sum_{s=1}^{s_{n}}\frac{\lambda
_{0}^{2}J_{n}V_{n}^{JK}}{\alpha _{n}^{\prime }\left( k\right) K_{n}}\left[
\sum_{l=\pm 1}\frac{K_{n+l}(\gamma _{1}r_{0})}{V_{n+l}^{JK}}\right]
_{k=k_{n,s}}^{2}.  \label{Ieig3}
\end{equation}

Let us compare the total energy flux with the radiated power $W^{\mathrm{(nm)%
}}$ expressed in terms of the work per unit time done by the radiation field
on the charge:%
\begin{equation}
W^{\mathrm{(nm)}}=-\int_{0}^{\infty }dr\int_{0}^{2\pi }d\phi \int_{-\infty
}^{\infty }dzr\,\mathbf{j}\cdot \mathbf{E}^{\mathrm{(r)}}.  \label{P}
\end{equation}%
For the current density (\ref{ji}), the component $E_{2}^{\mathrm{(r)}}(x)$
of the electric field is required, evaluated at $r=r_{0}$. Plugging the
representation (\ref{Xrad2}) for this component, by taking into account (\ref%
{E3se}) for $m=2$ and the first relation in (\ref{Rel12}), for the radiated
power on $n$-th harmonic we get%
\begin{equation}
W_{n}^{\mathrm{(nm)}}=\frac{q^{2}v^{2}}{\omega _{n}\left( \varepsilon
_{1}-\varepsilon _{0}\right) }\sum_{s=1}^{s_{n}}\frac{\lambda
_{0}^{2}J_{n}V_{n}^{JK}}{\alpha _{n}^{\prime }\left( k_{n,s}\right) K_{n}}%
\left[ \sum_{l=\pm 1}\frac{K_{n+l}(\gamma _{1}r_{0})}{V_{n+l}^{JK}}\right]
_{k=k_{n,s}}^{2}.  \label{Pn}
\end{equation}%
Comparing with (\ref{Ieig3}), we see that $W_{n}^{\mathrm{(nm)}}=2P_{n}^{%
\mathrm{(nm)}}$. Here the factor 2 takes into account the energy fluxes in
the regions $z>0$ and $z<0$. For real $\varepsilon _{j}$ the absorption is
absent and we could expect this energy balance. For a given angular velocity 
$\omega _{0}$, the energy fluxes and the radiated power are monotonically
decreasing functions of the rotation orbit radius $r_{0}$ and decay as $%
e^{-2\gamma _{1}r_{0}}$ for $\gamma _{1}r_{0}\gg 1$. An alternative
expression for the radiated power in the form of guided modes is given in 
\cite{Saha19}. The equivalence to a simpler representation (\ref{Pn}) can be
seen by using the first relation in (\ref{Rel12}). The energy fluxes and the
radiated power for guided modes generated by a point charge rotating inside
a cylindrical waveguide are investigated in \cite{Saha12}.

\begin{figure}[tbph]
\begin{center}
\epsfig{figure=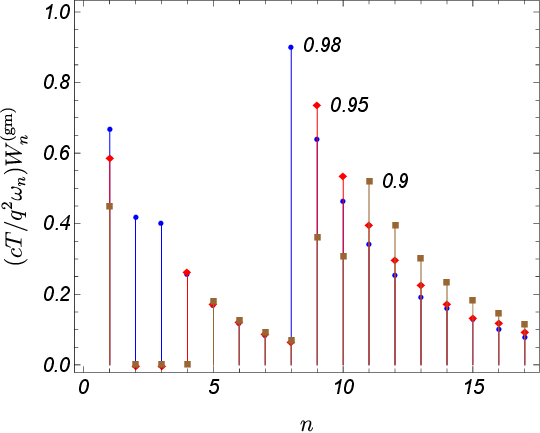,width=8cm,height=6.5cm}
\end{center}
\caption{The number of the radiated quanta per rotation period in the form
of guided modes (in units of $q^{2}/\hbar c$) versus of the radiation
harmonic. The numbers near the curves correspond to the values of $\protect%
\beta $. For the remaining parameters we have taken $\protect\varepsilon %
_{0}=1$, $\protect\varepsilon _{1}=3.8$, and $r_{1}/r_{0}=0.95$. }
\label{fig4}
\end{figure}

For numerical analysis of the radiated power in the form of guided modes, we
have considered an example with dielectric permittivities $\varepsilon
_{0}=3.8$ and $\varepsilon _{1}=1$, and with the parameters $\beta
=0.9,0.95,0.98$, $r_{1}/r_{0}=0.95$. For these values of the parameters and
for $\beta =0.98$, one has $s_{n}=1$ for $n=1,\ldots ,7$, $s_{n}=2$ for $%
n=8,\ldots ,13$, $s_{n}=3$ for $n=14,\ldots ,17$, and so on. In the case $%
\beta =0.95$ ($\beta =0.9$), the radiation is absent on the harmonics $n=2,3$
($n=2,3,4$), $s_{n}=1$ for $n=1,4,\ldots ,8$ ($n=1,5,\ldots ,8$), and $%
s_{n}=2$ for $n=9,\ldots ,17$. In Fig. \ref{fig4}, we have plotted the total
flux of the number of quanta for guided modes, radiated per rotation period,
versus the harmonic number. The numbers near the plot markers correspond to
the values of $\beta $. The corresponding values for $k_{n,s}$ lie in the
range (\ref{kguided}). As seen from Fig. \ref{fig4}, the radiation intensity
essentially increases with the appearance of a new mode $k_{n,s}$ for a
given $n$. This happens on the mode $n=8$ for $\beta =0.98$, and on the mode 
$n=9$ for $\beta =0.9,0.95$.

\section{Radiation of surface polaritons}

\label{sec:SP}

The formulas given above for the radiation on the normal modes of a
dielectric waveguide are valid for both guided and surface polaritonic
modes. For convenience of the further use, this section presents the
formulas for the characteristics of the radiated surface polaritons and
describes their radiation features. Various methods are used to excite
surface polaritons. These include prism and grating coupling, guided
photonic modes of waveguide, focused optical and electron beams (see \cite%
{Abaj10,Maie07,Enoc12,Stoc18,Han13,Gong14}).

For surface polaritons, the quantity $\lambda _{0}$ is purely imaginary and
the corresponding formulas for the radiation energy fluxes from a charge
rotating around a dielectric cylinder are obtained from the expressions
given above by the substitutions $\lambda _{0}=i\gamma _{0}$ and $%
J_{n}(\lambda _{0}r)=e^{in\pi /2}I_{n}(\lambda _{0}r)$. The energy flux
through the plane $z=\mathrm{const}$ in the region $r<r_{1}$ reads%
\begin{align}
P_{0n}^{\mathrm{(sp)}} &=\frac{q^{2}v^{2}}{4\omega _{n}\left( \varepsilon
_{1}-\varepsilon _{0}\right) }\sum_{s=1}^{s_{n}}\frac{k\gamma _{11}\gamma
_{01}^{2}I_{n}V_{n}^{IK}}{\alpha _{n}^{\prime 2}\left( k_{n,s}\right) }\left[
\sum_{l=\pm 1}\frac{K_{n+l}\left( \gamma _{1}r_{0}\right) }{lV_{n+l}^{IK}}%
\right] ^{2}  \notag \\
&\times \sum_{p=\pm 1}\frac{K_{n+p}}{\left( V_{n+p}^{IK}\right) ^{2}}\left[
\left( 1+\left( \frac{n+p}{\gamma _{01}}\right) ^{2}\right)
I_{n+p}^{2}-I_{n+p}^{\prime 2}\right] _{k=k_{n,s}},  \label{I0sp}
\end{align}%
where $I_{\nu }=I_{\nu }(\gamma _{01})$, $K_{\nu }=K_{\nu }(\gamma _{11})$, $%
\nu =n,n\pm 1$, and 
\begin{equation}
V_{n}^{IK}=\gamma _{11}I_{n}K_{n}^{\prime }-\gamma _{01}K_{n}I_{n}^{\prime
}=-\gamma _{01}I_{n+p}K_{n}-\gamma _{11}I_{n}K_{n+p},  \label{VnIK}
\end{equation}%
for $p=\pm 1$. Note that the latter function is always negative. For the
energy flux outside the cylinder, $r>r_{1}$, one obtains 
\begin{align}
P_{1n}^{\mathrm{(sp)}} &=\frac{q^{2}v^{2}}{4\omega _{n}\left( \varepsilon
_{0}-\varepsilon _{1}\right) }\sum_{s=1}^{s_{n}}\frac{k\gamma
_{01}^{3}I_{n}^{2}V_{n}^{IK}}{\alpha _{n}^{\prime 2}\left( k_{n,s}\right)
K_{n}}\left[ \sum_{l=\pm 1}\frac{K_{n+l}\left( \gamma _{1}r_{0}\right) }{%
lV_{n+l}^{IK}}\right] ^{2}  \notag \\
&\times \sum_{p=\pm 1}\frac{I_{n+p}}{\left( V_{n+p}^{IK}\right) ^{2}}\left[
K_{n+p}^{\prime 2}-\left( 1+\left( \frac{n+p}{\gamma _{11}}\right)
^{2}\right) K_{n+p}^{2}\right] _{k=k_{n,s}}.  \label{I1sp}
\end{align}%
The functions in the square brackets of (\ref{I0sp}) and (\ref{I1sp}) under
the sign of the summation over $p$ are obtained from the radial integrals $%
\int_{0}^{r_{1}}dr\,rI_{n+p}^{2}(\gamma _{0}r)$ and $\int_{r_{1}}^{\infty
}dr\,rK_{n+p}^{2}(\gamma _{0}r)$, and they are positive. Hence, the signs of
the energy fluxes are determined by the sign of the ratio $V_{n}^{IK}/\left(
\varepsilon _{1}-\varepsilon _{0}\right) $. By taking into account that $%
V_{n}^{IK}<0$, we conclude that 
\begin{equation}
\left( \varepsilon _{0}-\varepsilon _{1}\right) P_{1n}^{\mathrm{(sp)}%
}<0<\left( \varepsilon _{0}-\varepsilon _{1}\right) P_{0n}^{\mathrm{(sp)}}.
\label{I01spsign}
\end{equation}%
In particular, this shows that the energy fluxes for surface polaritons have
opposite signs inside and outside the cylinder. We recall that for the
presence of surface polaritons the permittivities $\varepsilon _{0}$ and $%
\varepsilon _{1}$ should have opposite signs. Combining this with (\ref%
{I01spsign}), we conclude that the energy flux is positive/negative in a
medium with positive/negative dielectric permittivity.

The total energy flux for radiated surface polaritons is expressed as%
\begin{equation}
P_{n}^{\mathrm{(sp)}}=\frac{q^{2}v^{2}}{2\omega _{n}\left( \varepsilon
_{1}-\varepsilon _{0}\right) }\sum_{s=1}^{s_{n}}\frac{\gamma
_{0}^{2}I_{n}V_{n}^{IK}}{\alpha _{n}^{\prime }\left( k_{n,s}\right) K_{n}}%
\left[ \sum_{l=\pm 1}\frac{K_{n+l}(\gamma _{1}r_{0})}{lV_{n+l}^{IK}}\right]
_{k=k_{n,s}}^{2}.  \label{Isp}
\end{equation}%
For the radiated power $W^{\mathrm{(sp)}}$ of the surface polaritons we have 
$W_{n}^{\mathrm{(sp)}}=2P_{n}^{\mathrm{(sp)}}$. The radiated power was
investigated in \cite{Kota19} by evaluating the work done by the radiation
field on a charged particle. The explicit expression of the derivative $%
\alpha _{n}^{\prime }\left( k_{n,s}\right) $ in the formulas (\ref{I0sp})-(%
\ref{Isp}) reads 
\begin{align}
\alpha _{n}^{\prime }\left( k\right) =&-\frac{r_{1}^{3}k}{2}\sum_{l=\pm
1}\left( \frac{I_{n}K_{n}}{V_{n+l}^{IK}}\right) ^{2}\left\{ \gamma _{1}\frac{%
K_{n+l}}{K_{n}}\left[ 1+\left( \frac{n+l}{\gamma _{01}}\right) ^{2}-\left( 
\frac{I_{n}^{\prime }}{I_{n}}+\frac{1}{\gamma _{01}}\right) ^{2}\right]
\right.  \notag \\
&\left. +\gamma _{0}\frac{I_{n+l}}{I_{n}}\left[ 1+\left( \frac{n+l}{\gamma
_{11}}\right) ^{2}-\left( \frac{K_{n}^{\prime }}{K_{n}}+\frac{1}{\gamma _{11}%
}\right) ^{2}\right] \right\} _{k=k_{n,s}}.  \label{alfnder2}
\end{align}%
The dispersion relation $\omega =\omega (k)$ for surface polaritons is
obtained from (\ref{NormMode3}). In the problem under consideration, the
values $k=k_{n,s}$ for the radiated surface polaritons are determined by the
intersection points of the curve $\omega =\omega (k)$ and the line $\omega
=\omega _{n}$ in the $(k,\omega )$-plane. An equivalent form of the
dispersion equation reads%
\begin{equation}
\alpha _{n}(k)=\frac{\varepsilon _{0}}{\varepsilon _{1}-\varepsilon _{0}}+%
\frac{1}{2}\sum_{l=\pm 1}\left( 1+\frac{\gamma _{1}I_{n+l}K_{n}}{\gamma
_{0}I_{n}K_{n+l}}\right) ^{-1}=0.  \label{DispSP}
\end{equation}%
For a given value of the combination $\beta _{1}=\beta \sqrt{\varepsilon _{1}%
}$, the dimensionless product $k_{n,s}r_{1}$ is a function of $n$, $%
r_{1}/r_{0}$, and $\varepsilon _{0}/\varepsilon _{1}$.

Let us consider the asymptotic properties of the radiated surface
polaritonic modes. Unlike the guided modes, for surface polaritons with a
given frequency $\omega _{n}$, the values of the projection $k$ of the wave
vector are limited only from below. In the analysis of the radiated surface
polariton modes it is convenient to introduce the dimensionless combination $%
z_{n}=kc/\omega _{n}$. This combination is the ratio of the light wavelength
of angular frequency $\omega _{n}$ in the vacuum to the wavelength of
surface polariton with the same frequency. For large values of $k$ one has $%
z_{n}\gg 1$ and the arguments of the modified Bessel functions are large. By
using the corresponding asymptotic formulas, in the leading order, the
dispersion equation is reduced to 
\begin{equation}
\alpha _{n}(k)\approx \frac{1}{\varepsilon _{1}/\varepsilon _{0}-1}+\frac{1}{%
\gamma _{1}/\gamma _{0}+1}=0,  \label{alfas1}
\end{equation}%
having the solution 
\begin{equation}
k^{2}\approx \frac{\omega _{n}^{2}}{c^{2}}\frac{\varepsilon _{0}\varepsilon
_{1}}{\varepsilon _{0}+\varepsilon _{1}}.  \label{klarge}
\end{equation}%
Note that the dispersion relation $\omega =\omega (k)$ for surface
polaritons on a planar boundary, separating media with dielectric
permittivities $\varepsilon _{0}$ and $\varepsilon _{1}$, is given by (\ref%
{klarge}) with $\omega _{n}$ replaced by $\omega $. Such an approximation by
the dispersion relation for a flat boundary was expected, since the limit
under consideration corresponds to small wavelengths, for which the
curvature of the cylinder is not essential. In particular, the relation (\ref%
{klarge}) shows that $\varepsilon _{0}\rightarrow -\varepsilon _{1}$ in the
limit $k\rightarrow \infty $.

The features caused by the curvature of the cylinder surface are important
for wavelengths of the order of the cylinder radius and larger. To be
specific, let us consider the case $\varepsilon _{0}<0<\varepsilon _{1}$.
With this choice, the minimal value of $k$ is determined from the condition $%
\lambda _{1}^{2}<0$ and $k\geq \omega _{n}\sqrt{\varepsilon _{1}}/c$. For
the values of $k$ close to this lower limit, assuming that $\gamma _{11}\ll
1 $, in (\ref{DispSP}) we replace the Macdonald function by its asymptotic
for small values of the argument. In addition, one has $\gamma _{01}\approx
(\omega _{n}r_{1}/c)\sqrt{|\varepsilon _{0}|+\varepsilon _{1}}$. For the
lowest radiated harmonic $n=1$, from the dispersion relation (\ref{DispSP})
we obtain%
\begin{equation}
\alpha _{1}(k)\approx \frac{1}{1-\varepsilon _{0}/\varepsilon _{1}}+\frac{%
I_{0}}{2\gamma _{01}I_{1}\ln \gamma _{11}}=0.  \label{alfas2}
\end{equation}%
From here it follows that the modes with $k$ close to the lower limit are
radiated in the spectral range where $\varepsilon _{0}/\varepsilon _{1}\ll
-1 $. The situation is completely different for the harmonics $n\geq 2$. In
this case, under the condition $\gamma _{11}\ll 1$, we find%
\begin{equation}
\alpha _{n}(k)\approx \frac{1}{2}\left( \frac{1+\varepsilon _{0}/\varepsilon
_{1}}{1-\varepsilon _{0}/\varepsilon _{1}}+\frac{I_{n}}{I_{n-2}}\right) =0.
\label{alfas3}
\end{equation}%
By taking into account that for $v_{c}|\varepsilon _{0}|/c\gg 1$ the
argument of the modified Bessel functions is large and using the
corresponding asymptotics, it can be shown that for modes $n>1$ and in the
region $|\varepsilon _{0}|\gg 1$ the equation (\ref{alfas3}) has no
solutions. For a given $n\geq 2$ one has a threshold $\varepsilon _{0n}$ for
the cylinder dielectric permittivity and the equation (\ref{DispSP}) has no
roots in the region $\varepsilon _{0}<\varepsilon _{0n}$. When $k$ tends to
its lower limit $\omega _{n}\sqrt{\varepsilon _{1}}/c$, the dielectric
permittivity $\varepsilon _{0}$ approaches to a finite value $\varepsilon
_{0n}^{(1)}$. With increasing $v_{c}/c$, the limiting value $\varepsilon
_{0n}^{(1)}$ tends to $\varepsilon _{0n}$.

\begin{figure}[tbph]
\begin{center}
\begin{tabular}{cc}
\epsfig{figure=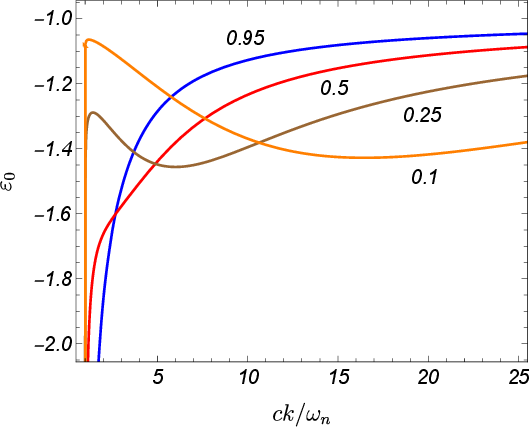,width=8cm,height=6.5cm} & \quad %
\epsfig{figure=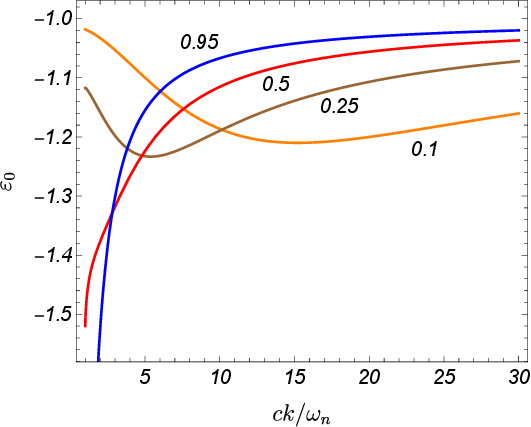,width=8cm,height=6.5cm}%
\end{tabular}%
\end{center}
\caption{The surface polaritonic modes of a dielectric cylinder with respect
to the ratio $z_{n}=kc/\protect\omega _{n}$. The numbers near the curves are
the values of $\protect\beta $ and the graphs are plotted for $n=1$ (left
panel) and $n=2$ (right panel). We have taken $\protect\varepsilon =1$ for
the exterior medium.}
\label{fig5}
\end{figure}

The features described above for the distribution of the surface polariton
modes are numerically illustrated in Fig. \ref{fig5}. In this analysis it is
convenient to consider (\ref{DispSP}) as an equation with respect to $%
\varepsilon _{0}$ for a given value of $z_{n}=kc/\omega _{n}$. This
dependence is plotted for $\varepsilon _{1}=1$, $r_{1}/r_{0}=0.95$, and for
different values of $\beta $ (the numbers near the curves). The left and
right panels correspond to the harmonics $n=1$ and $n=2$, respectively.
According to asymptotic analysis, in both cases we have $\varepsilon
_{0}\rightarrow -1$ in the limit $z_{n}\gg 1$. In the limit $%
z_{n}\rightarrow 1$, one has $\varepsilon _{0}\rightarrow -\infty $ for $n=1$
and $\varepsilon _{0}\rightarrow $ $\varepsilon _{0n}^{(1)}$ for $n=2$. The
limiting values are given by $\varepsilon _{0n}^{(1)}\approx
-1.02,-1.12,-1.52,-3.41$ for $\beta =0.1,0.25,0.5,0.95$, respectively. As
seen from the graphs, $\varepsilon _{0n}^{(1)}=\varepsilon _{0n}$ for $\beta
=0.5,0.95$. For $\beta =0.1,0.25$ one has $\varepsilon _{0n}\approx
-1.21,-1.23$. In the cases $\beta =0.5,0.95$ and for a given $\varepsilon
_{0}<-1$, there is a single root for $n=1$ and a single root in the range $%
\varepsilon _{0n}<\varepsilon _{0}<-1$ for $n=2$. For $\beta =0.1,0.25$,
depending on the value of $\varepsilon _{0}$, one can have one, two or three
roots.

\begin{figure}[tbph]
\begin{center}
\begin{tabular}{cc}
\epsfig{figure=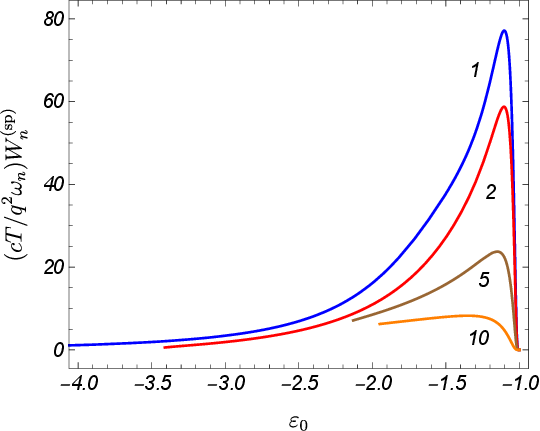,width=8cm,height=6.5cm} & \quad %
\epsfig{figure=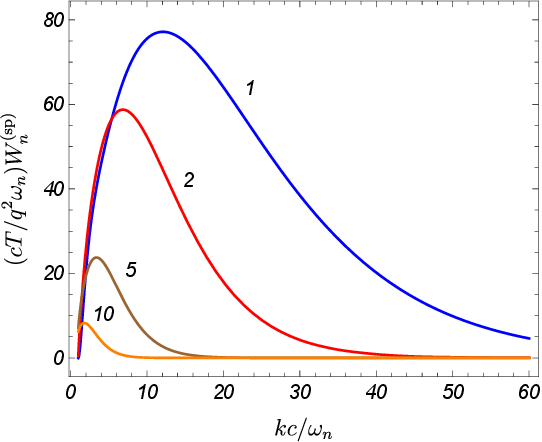,width=8cm,height=6.5cm}%
\end{tabular}%
\end{center}
\caption{The number of the quanta, radiated per rotation period, in the form
of surface polaritons (in units of $q^{2}/\hbar c$), as a function of the
cylinder dielectric permittivity (left panel) and as a function of $z_{n}=kc/%
\protect\omega _{n}$ (right panel). The numbers near the curves are the
values of the radiated harmonic $n$. For the parameters we have taken $%
\protect\varepsilon _{1}=1$, $r_{1}/r_{0}=0.95$, $\protect\beta =0.95$.}
\label{fig6}
\end{figure}

Figure \ref{fig6} presents the number of the radiated quanta for surface
polaritons per rotation period $T$ (in units of $q^{2}/\hbar c$) as a
function of $\varepsilon _{0}$ (left panel) and $z_{n}=kc/\omega _{n}$
(right panel), for $\varepsilon _{1}=1$, $r_{1}/r_{0}=0.95$, $\beta =0.95$.
The numbers near the curves present the values of $n$. The dependence $z_{n}$
on the dielectric permittivity $\varepsilon _{0}$ for $n=1,2$ is given in
Fig. \ref{fig5}. As seen, the number of the quanta radiated in the form of
surface polaritons can be essentially larger than the one for guided modes.

The above discussion is presented without specifying the dispersion of the
dielectric permittivites $\varepsilon _{j}=\varepsilon _{j}(\omega )$. The
allowed values of $k_{n,s}$ for the normal modes of the cylinder, being the
zeros of the function $\alpha _{n}(k)$, depend on specific dispersion law.
For instance, in the case corresponding to the data in Fig. \ref{fig5}, the
roots $k_{n,s}$ are obtained by solving the equation $\varepsilon
_{0}(\omega _{n})=h_{n}(kc/\omega _{n})$ with respect to $k$ for a given $%
\omega _{n}$. Here, the function $h_{n}(kc/\omega _{n})$ is presented by the
graphs in Fig. \ref{fig5} for $n=1$ and $n=2$. As an example of dispersion,
the Drude model can be considered with $\varepsilon _{0}(\omega )=1-\omega
_{p}^{2}/\omega ^{2}$ with $\omega _{p}$ being the plasma frequency.

We considered the radiation of a charged particle rotating around a cylinder
as an application of the GF with the components given in Section \ref%
{sec:GFwaveguide}. Another class of problems employing different GF
components is discussed in \cite{Saha20,Saha23,Saha24,Saha25}. These papers
study the radiation of point particles and annular beams moving parallel to
the axis of a dielectric cylinder.

\section{Conclusion}

\label{sec:Conc}

We reviewed the features of radiation processes in a cylindrically symmetric
medium. For given sources of the electromagnetic field, the response of the
medium is encoded in the Green function. In cylindrical coordinates and for
a medium with a dielectric function $\hat{\varepsilon}(r)$, the equation for
the GF is presented in the form (\ref{GDeq}) with matrix operators (\ref%
{Fcal}) and (\ref{Dcal}). In accordance with the problem's symmetry, it is
convenient to switch to the partial Fourier representation (\ref{GDFour}).
Given \ a current density $\mathbf{j}(r)$, the Fourier component of the
vector potential can be found using the relation (\ref{AiGDf}). Section \ref%
{sec:RecRel} considers the special case of a cylindrically symmetric medium
consisting of $N$ homogeneous cylindrical layers described by dielectric
function (\ref{epscyl}). The equation for the corresponding GF is
transformed to the form (\ref{EqGF2}) with a set of delta-type "potentials" $%
\hat{A}^{(s)}(r)$ located at separating boundaries and defined by (\ref%
{Akpot}). Presenting the equation for the GF in the form of
Lipmann-Schwinger equation, the relation (\ref{Gsrec2}) is derived fo
intermediate GFs $\hat{G}_{n}^{(s)}(r,r^{\prime })$ defined by (\ref{GFs}).
For $s=N$, this function coincides with the GF for $N$ layers and the
equation (\ref{Gsrec2}) provides a recurrence relation to find the function $%
\hat{G}_{n}^{(N)}(r,r^{\prime })$ having the GF $\hat{G}_{n}^{(N-1)}(r,r^{%
\prime })$ for $N-1$ layers.

Using the recurrence relation (\ref{Gsrec2}), we reduce the problem to the
determining the GF $G_{n}^{(0)}\left( r,r^{\prime }\right) $ being the
solution of the equation (\ref{GD0eq}). In Section \ref{sec:G0}, we
diagonalized this equation by transformation (\ref{Trans1}). The diagonal
components of the transformed GF are found by solving the corresponding
equation in separate layers and imposing matching conditions at the
interfaces between them. By imposing the regularity condition on the
symmetry axis and the radiation condition in the exterior medium, one gets $%
2N+2$ equations for the $2N+2$ coefficients. They are expressed in terms of
the functions (\ref{VFG}) with $F_{n}(y)$ and $G_{n}(y)$ being the Bessel
and Hunkel functions, respectively. The initial GF\ $G_{n}^{(0)}\left(
r,r^{\prime }\right) $ is obtained via inverse transformation using the
matrix (\ref{Mtrans}). This summarizes the general procedure for finding of
the GF\ $\hat{G}_{n}(r,r^{\prime })=\hat{G}_{n}^{(N)}(r,r^{\prime })$.

As an application of the general setup, Section \ref{sec:GFwaveguide}
considers a cylindrical waveguide with dielectric function $\varepsilon
_{0}(\omega )$ surrounded by a homogeneous medium with permittivity $%
\varepsilon _{1}(\omega )$. For convenience of the presentation of formulas,
we define the function $\hat{G}^{\mathrm{(c)}}(r,r^{\prime })$ as given in (%
\ref{Gc}). For points inside the cylinder, this function describes the
effects of the exterior medium (the effects of nonzero difference $%
\varepsilon _{1}-\varepsilon _{0}$), and for points in the exterior medium,
it gives the contribution induced by the cylinder. All the components of the
matrix $\hat{G}^{\mathrm{(c)}}(r,r^{\prime })$ are found. The components $%
G_{2l}^{\mathrm{(c)}}(r,r^{\prime })$, with $l=1,2$, become zero. The
non-zero components in the interior and exterior regions are given by (\ref%
{Glmi}), (\ref{G33i}), (\ref{Glme}), and (\ref{G33e}). Knowledge of these
components enables investigation of the radiation fields for the general
case of a radiating source $\mathbf{j}(r)$. Section \ref{sec:Features},
discusses general features for three different types of radiation. These
correspond to the radiation at large distances from the cylinder, radiation
in the form of guided modes of the cylinder, and radiation in the form of
surface polaritons localized near the cylinder surface. The latter type of
radiation is present in the spectral range where the dielectric
permittivities of the neighboring media have opposite signs. Conditions for
the appearance of strong peaks in the angular distribution of the radiation
propagating at large distances from the cylinder are specified. An equation
is derived for the location of these peaks.

In the second part of the paper, a point charge $q$ that rotates around a
cylinder with constant velocity $v$ is considered as a radiation source. In
Section \ref{sec: RadInf}, the expressions for the scalar and vector
potentials, and for the electric and magnetic fields inside and outside the
cylinder are presented. These fields are used to investigate the
spectral-angular density of the radiation intensity. For a given harmonic $n$%
, the angular density of the radiation intensity at large distances from the
cylinder is expressed as (\ref{I3}), with the functions $W_{n}^{(p)}$
defined in (\ref{Wpn}). Analyzing the presence of narrow peaks in the
particular case of the radiation source under consideration confirms the
conditions specified for the general case. The necessary conditions $%
\varepsilon _{0}>\varepsilon _{1}$ and $v_{c}\sqrt{\varepsilon _{0}}>c$ are
required for those peaks, where $v_{c}$ is the projection of the particle
velocity onto the cylinder surface. Under these conditions, the peaks are
located in the angular region (\ref{peakscond3}). A numerical analysis of
the radiation intensity shows that the results of the general analytic
estimates of the peak positions coincide with the data obtained by numerical
calculations with high accuracy.

Section \ref{sec:RadGuid} studies the electromagnetic fields and energy
fluxes of radiation from a circulating charge in the form of the waveguide
normal modes. These parts of the total field originate from the poles of the
GF, which are the zeros of the function $\alpha _{n}(k)$. The rules for
specifying the integration contour near these poles are dictated by
introducing a small imaginary part to the dielectric permittivity. The
radiation fields have the structure (\ref{Xrad2}), where the radial
functions for the magnetic and electric fields are given by the expressions (%
\ref{H3s}), (\ref{E3s}) inside the cylinder and by (\ref{H3se}), (\ref{E3se}%
) in the region $r>r_{1}$. As an energetic characteristic of the radiated
waves, the energy flux through a plane perpendicular to the cylinder's axis
is considered. For guided modes, the fluxes in the interior and exterior
regions are given by the expressions (\ref{I0eig2}) and (\ref{I1eig2}),
respectively. A simpler expression (\ref{Ieig3}) is obtained for the total
energy flux by combining the fluxes inside and outside the cylinder. The
radiated power, being the work done by the radiation field on the charge per
unit time, has been shown to be equal to twice the total energy flux in the
region $z>0$. The features of distribution of the radiated surface
polaritonic modes and the corresponding energy fluxes in the exterior and
interior regions are discussed in Section \ref{sec:SP}. The fluxes are
positive/negative in a medium with positive/negative permittivity. The total
energy flux and the radiated power are also studied. It has been
demonstrated that the number of quanta radiated in the form of surface
polaritons can be significantly greater than the corresponding quantity for
guided modes.

\section*{Acknowledgments}

A.A.S. was supported by the grant No. 21AG-1C047 of the Higher Education and
Science Committee of the Ministry of Education, Science, Culture and Sport
RA. L.Sh.G. and H.F.Kh. were supported by the grant No. 21AG-1C069 of the
Higher Education and Science Committee of the Ministry of Education,
Science, Culture and Sport RA.

\appendix

\section{Equations for the vector and scalar potentials in the time domain}

\label{sec:App}

The Maxwell equations in a nonmagnetic medium read (see, e.g, \cite{Jack99}
for a more general case of magnetic medium)%
\begin{align}
\mathbf{\nabla }\cdot \mathbf{D} =&4\pi \rho ,\;\mathbf{\nabla }\times 
\mathbf{E=-}\frac{1}{c}\partial _{t}\mathbf{B},  \notag \\
\mathbf{\nabla }\cdot \mathbf{B} =&0,\;\mathbf{\nabla }\times \mathbf{B}=%
\frac{4\pi }{c}\mathbf{j}+\frac{1}{c}\partial _{t}\mathbf{D}.\;  \label{Meq}
\end{align}%
In terms of the spectral components of the fields, defined by 
\begin{equation}
f(t,\mathbf{r})=\int_{-\infty }^{+\infty }d\omega \,\,f(\omega ,\mathbf{r}%
)e^{-i\omega t},\;f_{\omega }=f(\omega ,\mathbf{r}),  \label{Spec}
\end{equation}%
these equations become%
\begin{align}
\mathbf{\nabla }\cdot \mathbf{D}_{\omega } =&4\pi \rho _{\omega },\;\mathbf{%
\nabla }\times \mathbf{E}_{\omega }=\frac{i\omega }{c}\mathbf{B}_{\omega }, 
\notag \\
\mathbf{\nabla }\cdot \mathbf{B}_{\omega } =&0,\;\mathbf{\nabla }\times 
\mathbf{B}_{\omega }=\frac{4\pi }{c}\mathbf{j}_{\omega }-\frac{i\omega }{c}%
\mathbf{D}_{\omega }.\;  \label{Meqom}
\end{align}

For an isotropic, time-invariant medium without spatial dispersion, we have
the constitutive relation%
\begin{equation}
\mathbf{D}(\omega ,\mathbf{r})=\varepsilon _{\omega }\mathbf{E}(\omega ,%
\mathbf{r}),\;\varepsilon _{\omega }=\varepsilon \left( \omega ,\mathbf{r}%
\right) ,  \label{DEom}
\end{equation}%
and the equations (\ref{Meqom}) take the form%
\begin{align}
& \varepsilon _{\omega }\mathbf{\nabla }\cdot \mathbf{E}_{\omega }+\left( 
\mathbf{\nabla }\varepsilon _{\omega }\right) \cdot \mathbf{E}_{\omega
}=4\pi \rho _{\omega },\;\mathbf{\nabla }\times \mathbf{E}_{\omega }=\frac{%
i\omega }{c}\mathbf{B}_{\omega },  \notag \\
& \mathbf{\nabla }\cdot \mathbf{B}_{\omega }=0,\;\mathbf{\nabla }\times 
\mathbf{B}_{\omega }=\frac{4\pi }{c}\mathbf{j}_{\omega }-\frac{i\omega }{c}%
\varepsilon _{\omega }\mathbf{E}_{\omega }.\;  \label{Meqom2}
\end{align}%
Next, we introduce the spectral components of the scalar and vector
potentials in accordance with%
\begin{equation}
\mathbf{E}_{\omega }=\frac{i\omega }{c}\mathbf{A}_{\omega }-\mathbf{\nabla }%
\varphi _{\omega },\;\mathbf{B}_{\omega }=\mathbf{\nabla }\times \mathbf{A}%
_{\omega }.  \label{Potent}
\end{equation}%
Imposing the Lorentz gauge condition $\mathbf{\nabla }\cdot \mathbf{A}%
_{\omega }=i\omega \varepsilon _{\omega }\varphi _{\omega }/c$, the
equations for those components are written as (the corresponding equations
in a medium with nontrivial magnetic properties can be found in \cite{Ghar83}%
)%
\begin{align}
& \Delta \mathbf{A}_{\omega }+\varepsilon _{\omega }\frac{\omega ^{2}}{c^{2}}%
\mathbf{A}_{\omega }-\frac{\mathbf{\nabla }\varepsilon _{\omega }}{%
\varepsilon _{\omega }}\mathbf{\nabla }\cdot \mathbf{A}_{\omega }=-\frac{%
4\pi }{c}\mathbf{j}_{\omega },  \notag \\
& \mathbf{\Delta }\varphi _{\omega }+\varepsilon _{\omega }\frac{\omega ^{2}%
}{c^{2}}\varphi _{\omega }-\frac{\mathbf{\nabla }\varepsilon _{\omega }}{%
\varepsilon _{\omega }}\cdot \left( \frac{i\omega }{c}\mathbf{A}_{\omega }-%
\mathbf{\nabla }\varphi _{\omega }\right) =-\frac{4\pi }{\varepsilon
_{\omega }}\rho _{\omega }.  \label{Meqom3}
\end{align}

Introducing the function 
\begin{equation}
\varepsilon \left( t,\mathbf{r}\right) =\int_{-\infty }^{+\infty }d\omega
\,\varepsilon \left( \omega ,\mathbf{r}\right) e^{-i\omega t},  \label{epst}
\end{equation}%
the relationship between electric displacement and the electric field in
time domain is expressed in the well-known integral form%
\begin{equation}
\mathbf{D}(t,\mathbf{r})=\int_{-\infty }^{t}dt^{\prime }\,\varepsilon \left(
t-t^{\prime },\mathbf{r}\right) \mathbf{E}(t^{\prime },\mathbf{r}).
\label{DEt}
\end{equation}%
When deriving this relation from the corresponding formula for the spectral
components, we took into account that the function $\varepsilon \left(
\omega ,\mathbf{r}\right) $ is analytic in the half-plane $\mathrm{Im}%
\,\omega >0$ (see, e.g., \cite{Jack99,Land84}) and $\varepsilon \left(
t-t^{\prime },\mathbf{r}\right) =0$ for $t^{\prime }>t$. Let us introduce
the operator $\hat{\varepsilon}=\hat{\varepsilon}(\mathbf{r})$ the action of
which, for a given function $g(z)$, is defined as follows (note that $\hat{%
\varepsilon}$ is different from (\ref{epst})):%
\begin{equation}
g(\hat{\varepsilon})f(t,\mathbf{r})=\int_{-\infty }^{+\infty }d\omega
\,g(\varepsilon (\omega ,\mathbf{r}))f(\omega ,\mathbf{r})e^{i\omega t}.
\label{geps}
\end{equation}%
According to this definition, the time-domain equations corresponding to (%
\ref{Meqom3}) are written in the form (\ref{EqA}) and (\ref{Eqxi}).

Note that we could have originally conducted the above discussion in the
terms of spectral components, based on the equation (\ref{Meqom3}) for the
vector potential. The equation for the spectral components $A_{l}(\omega ,%
\mathbf{r})$ is obtained from (\ref{EqA2}) by the replacements $\hat{%
\varepsilon}\rightarrow \varepsilon (r)=\varepsilon \left( \omega ,r\right) $
and $\partial _{t}^{2}\rightarrow -\omega ^{2}$. The properties of the
medium do not depend on time and the GF depends on $t$ and $t^{\prime }$
through $t-t^{\prime }$. The corresponding spectral component is defined as%
\begin{equation}
G_{il}(x,x^{\prime })=\int_{-\infty }^{+\infty }d\omega \,G_{il}(\omega ,%
\mathbf{r},\mathbf{r}^{\prime })e^{-i\omega (t-t^{\prime })},  \label{Gilom}
\end{equation}%
and the vector potential is obtained from the relation 
\begin{equation}
A_{i}(\omega ,\mathbf{r})=-\frac{1}{\pi c}\,\int d\mathbf{r}^{\prime
}\,G_{il}(\omega ,\mathbf{r},\mathbf{r}^{\prime })\,j_{l}(\omega ,\mathbf{r}%
^{\prime }).  \label{AiG}
\end{equation}%
If we expand the Fourier series over the angular and axial coordinates, the
rest of the discussion is the same as in the main text.

\end{document}